\documentclass[a4paper,11pt]{article}
\usepackage[utf8]{inputenc}
\usepackage[english]{babel}
\usepackage{braket}
\usepackage{xspace}
\usepackage{bbm}
\usepackage{mathdots}
\usepackage{stackrel}
\usepackage[dvipsnames]{xcolor}
\usepackage{appendix}
\usepackage{hyperref}
\hypersetup{
 colorlinks=true,
 linkcolor=blue,
 anchorcolor = blue,
 citecolor = blue,
 filecolor = blue,
 urlcolor = blue
}
\usepackage{mathtools}
\usepackage{bbm}
\usepackage{comment}
\usepackage{subfigure}
\usepackage[T1]{fontenc}               
\usepackage[left=3cm,
            right=2.5cm,
            top=2.5cm,
            bottom=3cm
            ]{geometry}                
\usepackage{graphicx}
\usepackage{mathrsfs}
\usepackage{appendix}
\usepackage{fancyhdr}
\usepackage{amsmath,                   
            amssymb,                   
            amsthm}                    

\usepackage{setspace}
\usepackage{cite}
\usepackage{cancel}
\usepackage{bm}
\usepackage[margin=30pt, bf, font=small, center, justification=justified]{caption}[2004/07/16]

\usepackage{setspace} 

\usepackage{tikz}
\usepackage{bbm}

\newcommand{\bcirc}{\vcenter{\hbox{\scalebox{0.7}{$\bigcirc$}}}}
\newcommand{\bsquare}{\vcenter{\hbox{\scalebox{0.89}{$\square$}}}}

\title{\bf     Universality of Measurement-Induced Criticality under Symmetry-Breaking Measurements
}

\author{Angelo Russotto$^1$, Filiberto Ares$^1$, Pasquale Calabrese$^1$}

\date{}

\begin{document} 

\maketitle

{\small
\vspace{-5mm}  \ \\
{$^{1}$}  SISSA and INFN, via Bonomea 265, 34136 Trieste, Italy\\[-0.1cm]
\medskip
}

\begin{abstract}
We study the critical properties of random quantum circuits with a $U(1)$ symmetry subject to local projective measurements that explicitly break this symmetry. We find that, at the measurement-induced phase transition, symmetry-breaking measurements act as a relevant perturbation at large scales, leading to the same universal critical properties as the corresponding monitored random circuit with non-symmetric unitary dynamics. In particular, we consider monitored $U(1)$-symmetric Haar-random circuits in the limit of large local Hilbert-space dimension, where the trajectory-averaged entanglement entropy can be exactly obtained in terms of a classical statistical mechanics model. In this model, the charge associated with the conservation law follows a symmetric simple exclusion process, in which symmetry-breaking measurements correspond to disordered defects that create and destroy charges.
We prove that the charge correlation length remains finite for any measurement rate, ruling out a charge-sharpening transition, in contrast to the case of symmetry-preserving measurements. We further support our predictions at finite local Hilbert-space dimension through numerical finite-size scaling analyses of the entanglement transition in monitored $U(1)$-symmetric Haar and stabilizer random circuits.
\end{abstract}

\tableofcontents
\section{Introduction}

The repeated measurement of a non-equilibrium quantum many-body system can dramatically reshape its dynamics~\cite{fisher-rev, pv-22}. While unitary evolution preserves quantum coherence and generates entanglement, measurements break unitarity by projecting quantum states onto measurement outcomes, thereby destroying entanglement. The competition between these two processes gives rise to new classes of dynamical phases with no equilibrium counterpart. Understanding the nature of these phases and their universal properties has become a central challenge in the study of quantum many-body systems.

In closed non-equilibrium many-body systems, the entanglement entropy of a subsystem typically grows in time and eventually saturates to a value proportional to the subsystem size~\cite{cc-05, ac-17, nrvh-17, jhn-18}. In contrast, when local projective measurements are interspersed during the evolution and their rate is sufficiently large, the system can enter an area-law phase in which the entanglement saturates to a value independent of the subsystem size~\cite{srn19, lcf-18, cnpd-19}. These two regimes are separated by a critical point defining a measurement-induced phase transition (MIPT) in the steady state of individual quantum trajectories. These transitions have been extensively explored in a wide range of monitored quantum systems~\cite{ctl-19,  srs-20, abd-21, bmad-21, jlcsz-21, msks-22, kmr23, stfd-22, fava23, poboiko-23, fava24, ppgm-25, pgm-24,difresco24, tbfds-21,delmonte25, sfs25,kts25}, revealing universal properties and scaling behavior akin to those of equilibrium phase transitions. They have also been experimentally realized in different quantum simulators~\cite{noel22,koh23,google23, alpvgp-24, kstgfmm-25}.

A paradigmatic setting for studying MIPTs is provided by random quantum circuits~\cite{fisher-rev, pv-22, vasseur-lectures, skinner-lectures}. These models replace the microscopic details of a specific Hamiltonian by randomly sampled unitary gates, thereby focusing on the generic features of the quantum evolution. In the presence of local measurements, this framework captures the essential competition between unitary entanglement growth and measurement-induced disentanglement in its most minimal form. Despite their simplicity, these models exhibit entanglement phases and non-trivial critical behavior, making them a natural framework for investigating universality of MIPTs. Different geometries and gate ensembles have been considered in the literature~\cite{srn19, lcf-18, cnpd-19, lcf-19, gh20, zabalo20, ilc-20, tfd-20, lab-21, nrsr-21, igghk-21, srs-19, lab-21-2, zabalo22, bbcay-22,sierant22, jsbl-23, feng25, ha24, deluca25, khanna26}, most notably Haar-random gates, drawn uniformly from the unitary group, and Clifford gates, drawn uniformly from the Clifford group. In particular, Haar-random circuits are maximally generic and capture universal features of chaotic quantum dynamics. For these ensembles of gates, using the replica trick, the average entanglement entropy over gate realizations, measurement locations, and quantum trajectories can be exactly mapped to two-dimensional classical statistical models~\cite{jian20, bca-20, li24}. These models are analytically tractable in the large local Hilbert space limit, where they reduce to percolation. At finite local Hilbert-space dimension, however, they are not amenable to analytical treatment and one must rely on numerical methods~\cite{srn19, gh20, bca-20, lcf-19, zabalo20, zabalo22,sierant22}. In this scenario, stabilizer circuits provide a particularly powerful framework, as they are efficiently classically simulable and allow for large-scale numerical studies of entanglement entropy.

The presence of symmetries can qualitatively alter the nature of the MIPT and its associated universality class, and enrich the phase diagram~\cite{bca-21, hc-22, magpvy-23, dmgk-24, st23,ll25}. A prominent example
is given by $U(1)$-symmetric Haar-random monitored circuits, where both unitary gates and local projective measurements respect a global charge conservation~\cite{vasseur22}. In this setting, entanglement dynamics is intertwined with charge transport, giving rise to a charge sharpening transition within the volume-law phase in the steady state~\cite{vasseur22, barratt22, barratt22-2, oshima-23, chakraborty24, ik-24, gfjl-25, nj25, gmv26}: above a finite measurement rate, charge fluctuations become short-ranged and decay exponentially, while below the transition they remain delocalized with algebraic decay. In the large local Hilbert-space limit, the MIPT can again be mapped to a classical statistical model, but with additional degrees of freedom associated with the conserved charge. Despite this extra structure, the entanglement transition remains governed by percolation as in the absence of symmetries. At finite local Hilbert-space dimension, however, numerical studies have found that some universal properties of the MIPT differ from those of non-symmetric circuits~\cite{vasseur22}, indicating that the presence of conservation laws can modify the critical behavior.

\begin{figure}[t]
\includegraphics[width=\textwidth]{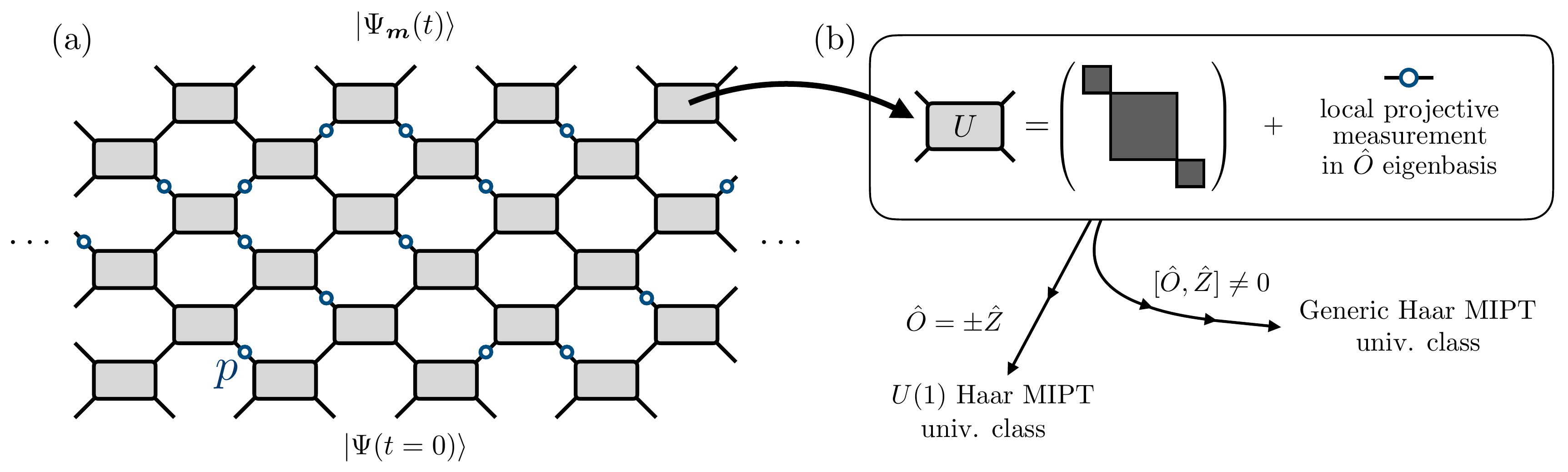}
\centering
\caption{(a) Sketch of the monitored random quantum circuit model considered in this paper. It has a brickwork structure in which each two-qubit gate $U$ conserves the magnetization $\hat{Z_j} + \hat{Z}_{j+1}$. Therefore, the gates have a block-diagonal structure in which the blocks, corresponding to fixed magnetization sectors, are independently drawn from the Haar unitary ensemble, as illustrated in the box. After a layer of unitary gates, the sites are projectively measured with probability $p$ in the eigenbasis of a local operator $\hat{O}$. The sites measured during the time evolution are represented by blue dots on the links between unitary gates. The system is initialized in the pure state $\ket{\Psi(0)}$, and its state at time $t$, $\ket{\Psi_{\bm{m}}(t)}$, depends on the set of measurement outcomes $\bm{m}$.
(b) Schematic summary of the main result of this paper. Local projective measurements $\hat{O}$ that break the symmetry of the unitary evolution act as a relevant perturbation, driving the system at large scales to the critical behavior of a generic monitored Haar random circuit with no symmetries in the unitary evolution. If $\hat{O}$ respects the symmetry, the system instead flows to a different universality class, as shown in Ref.~\cite{vasseur22}.}
\label{fig:sketch} 
\end{figure}

In this work, we investigate the interplay between symmetry and measurements from a complementary perspective. We consider 
the same $U(1)$-symmetric Haar-random circuit as in Ref.~\cite{vasseur22}, described in Fig.~\ref{fig:sketch}, but we monitor its dynamics with local projective measurements that explicitly break the conservation law. This setup allows us to address a fundamental question: whether symmetry-breaking measurements constitute a relevant perturbation for the measurement-induced critical point. We show that they do. Although the unitary dynamics remains strictly $U(1)$ symmetric, symmetry-breaking measurements drive the system to the same universality class as the MIPT in generic monitored Haar-random circuits without symmetries~\cite{srn19}. We support this conclusion analytically in the large local Hilbert-space limit, through the mapping to a statistical mechanics model, and numerically at finite local Hilbert-space dimension using both Haar-random and Clifford circuits. In the large local Hilbert-space limit, by mapping the charge sector to a symmetric exclusion process with disordered defects, we further show that charge fluctuations are characterized by a finite correlation length at any measurement rate, implying the absence of a charge sharpening transition, in contrast to symmetry-preserving measurements.

The paper is organized as follows.
In Sec.~\ref{sec:model} we introduce the monitor $U(1)$-symmetric Haar-random circuit that we study. In Sec.~\ref{sec:statmech}, we analytically analyze this circuit in the limit of large local Hilbert space dimension by mapping it to a classical statistical model. In Sec.~\ref{sec:qubitnum}, we present numerical results for the same circuit at finite local Hilbert space dimension. In Sec.~\ref{sec:stab}, we test our predictions on a stabilizer monitored circuit model, allowing for large scale numerical tests. Finally, Sec.~\ref{sec:Concl}, we draw our conclusions and provide possible outlooks for future research. In Appendix~\ref{app:sm}, we provide additional details on the statistical model that emerges in the large local Hilbert-space dimension limit. In Appendix~\ref{app:rndbas}, we discuss the statistical model associated with measurements performed in independently chosen random bases. Finally, in Appendix~\ref{app:num}, we describe the numerical methods used in Secs.~\ref{sec:qubitnum} and \ref{sec:stab}.

\section{Monitored Haar-random circuit model}
\label{sec:model}

In this section, we introduce the monitored circuit analyzed in the next two sections. The key feature of the setup is a unitary time evolution that respects a global $U(1)$ symmetry, together with a space-time-homogeneous monitoring protocol that explicitly breaks this symmetry.

Specifically, following Refs.~\cite{kvh-18, vasseur22}, we consider a chain of 
$L$ sites, each described by the tensor product of a qubit and a qudit, so that the 
local Hilbert space is $\mathcal{H}_1 = \mathbb{C}^2 \otimes \mathbb{C}^d$. The dynamics is generated by two-site random unitary gates applied in a brickwall pattern, as illustrated in Fig.~\ref{fig:sketch}. Each 
gate $U$ is drawn independently from the Haar ensemble of $4d^2 \times 4d^2$ 
unitary matrices commuting with the total qubit $z$-magnetization,
\begin{equation}
\label{eq:Qz}
\mathcal{Q} = \sum_{j=1}^L \,Z_j\, \otimes \,\mathbb{I}_j,
\end{equation}
where $Z_j$ denotes the Pauli-$Z$ operator acting on the qubit at site $j$, and $\mathbb{I}_j$ is the identity on the corresponding qudit. This symmetry implies that the gates can be decomposed into Haar-random unitaries within each sector of the conserved charge~\eqref{eq:Qz},
\begin{equation}
\label{eq:gatedecomp}
    U = \sum_q U_q  \Pi_q,
\end{equation}
with $q = 0, \pm 2$ labeling the charge sectors of the two qubits on which the gate acts, and $\Pi_q$ the corresponding  projectors. The dimensions of these sectors are $d^2$ for $q = \pm 2$ and $2d^2$ for $q = 0$. The matrices $U_q$ act non-trivially only within the charge-$q$ sector, where they are Haar-random unitaries.

After the application of a full layer (on either the even or odd bonds of the chain) of random unitary gates, each local degree of freedom is independently projectively measured with probability $p$ (see the sketch in Fig.~\ref{fig:sketch}). 
Both the qubit and the qudit are measured. The qubit is measured in the eigenbasis of the $x$-component of the magnetization, $X_j$, thereby explicitly breaking the conservation of the charge in Eq.~\eqref{eq:Qz}. This contrasts with Ref.~\cite{vasseur22}, where measurements are performed in the $Z_j$ eigenbasis, which preserves the charge conserved by the dynamics.  The qudit is measured in a fixed basis, but this choice is irrelevant since the dynamics in the qudit sector is fully Haar random and does not respect any symmetry.

For $d = 1$, the system corresponds to a one-dimensional chain of qubits in which the measured operator at each site is simply the Pauli matrix $X_j$. This case is analyzed numerically in Sec.~\ref{sec:qubitnum}. In the following section, we instead consider the limit $d \to \infty$, which can be treated analytically via a mapping of the circuit to a classical statistical model.

\section{Large-$d$ statistical-mechanics model}
\label{sec:statmech}

In this section, we study the classical statistical model emerging in the limit of large local Hilbert space dimension $d \to \infty$.
We first review the already known general formalism and then apply it to the model defined in Sec.~\ref{sec:model}. 
Analogous mappings have been discussed for Haar-random circuits both without~\cite{nvh-18, zn-19, krps-18, rpk-18, mcculloch23} and with monitoring~\cite{bca-20, jian20}, in particular for $U(1)$-symmetric circuits like ours but with projective measurements that respect the symmetry~\cite{vasseur22}.

\subsection{General formalism}

Let us consider a system initialized in the pure state
$\ket{\psi_0}$, and evolve it up to time $t$ under a dynamics involving both unitary gates and projective local measurements.  We first specify the set of measurement locations  $\bm{X}$, namely the spacetime positions at which measurements are performed up to time $t$, as indicated by circles in Fig.~\ref{fig:sketch}~(a). For this fixed measurement configuration,  the quantum trajectory of the system evolution is characterized by the collection of measurement outcomes and random gate realizations, which we collectively denote by $\bm{m}$.

At time $t$, the state of the full system for a given quantum trajectory $\bm{m}$ is described by the normalized density matrix 
\begin{equation}\label{eq:rho_m}
    \rho_{\bm{m}} = \frac{\check \rho_{\bm{m}}}{\text{Tr}[\check \rho_{\bm{m}}]},
\end{equation}
where the denominator is the Born probability for the trajectory $\bm{m}$, $p_{\bm{m}} = \text{Tr}[\check \rho_{\bm{m}}]$. The unnormalized density matrix $\check \rho_{\bm{m}}$ is obtained from the Kraus operator $K_{\bm{m}}$ of the monitored circuit, which encodes the evolution of the state conditioned on a given sequence of measurement outcomes, as $\check \rho_{\bm{m}} = K_{\bm{m}} \ket{\psi_0} \bra{\psi_0}K_{\bm{m}}^{\dagger}$.

Our goal is to study the entanglement entropy of a subsystem $A$, consisting of contiguous sites of the chain. For a given quantum trajectory $\bm{m}$ at time $t$, the state of subsystem $A$ is obtained by tracing out in Eq.~\eqref{eq:rho_m} the complementary subsystem $B$, yielding the reduced density matrix $\rho_{\bm{m}, A}={\rm Tr}_B(\rho_{\bm{m}})$. The R\'enyi-$n$ entropy of $A$ is then defined as
\begin{equation}
 S_n( \rho_{\bm{m}}) = \frac{1}{1-n} \log \text{Tr}[\rho_{\bm{m}, A}^{n}]. 
 \end{equation}
Using the standard approach~\cite{zzf-00, cc-04}, it is useful to write the R\'enyi entropies as a trace over the replicated full-system Hilbert space $\mathcal{H}^{\otimes n}$,
\begin{equation}
\label{eq:Sn}
S_n( \rho_{\bm{m}})= \frac{1}{1-n} \left(\log \text{Tr}[\check \rho_{\bm{m}}^{\otimes n} \,\mathcal{T}_{n,A}] - \log \text{Tr}[\check \rho_{\bm{m}}^{\otimes n}]\right),
\end{equation}
where $\mathcal{T}_{n,A}$ is the operator that cyclically permutes the $n$ replicas of the subsystem $A$. 

Let us now denote by $\mathbb{E}_{\bm{m}}[\cdot]$ the average over quantum trajectories at fixed measurement locations $\bm{X}$, i.e. the average over all the measurement outcomes and realizations of the random gates. This average is given by 
\begin{equation}\label{eq:traj_av}
    \mathbb{E}_{\bm{m}}[S_n] = \sum_{\bm{m}} \langle \, p_{\bm{m}}\,  S_n( \rho_{\bm{m}}) \, \rangle,
\end{equation}
where we denoted as $\langle\cdot\rangle $ the expectation value over all independently drawn unitary gates.
To avoid averaging the logarithm in Eq.~\eqref{eq:Sn}, one can use the replica trick~\cite{bca-20, jian20}, obtaining
\begin{equation}
\label{eq:replica}
S_n(\rho_{\bm{m}}) = \lim_{k\to 0} \frac{1}{k(1-n)} \left( \text{Tr}[\check \rho_{\bm{m}}^{\otimes kn} \,\mathcal{T}_{n,A}^{\otimes k}] - \text{Tr}[\check \rho_{\bm{m}}^{\otimes kn}]\right).
\end{equation}
We can further absorb the Born probability $p_{\bm{m}}$ appearing in Eq.~\eqref{eq:traj_av} into this expression by introducing one additional replica, such that the average over quantum trajectories is rewritten as
\begin{equation}
\mathbb{E}_{\bm{m}}[S_n] = \sum_{\bm{m}} \langle\tilde{S}_n(\rho_{\bm{m}})\rangle,
\end{equation}
where now
\begin{equation}
\label{eq:snZ}
    \tilde{S}_n(\rho_{\bm{m}}) =\lim_{k\to0}\frac{1}{k(1-n)} (Z_A(\bm{m})- Z_0(\bm{m})),
\end{equation}
with
\begin{equation}
\label{eq:Zs}
\begin{cases}
    Z_A(\bm{m}) = \langle \text{Tr}[\check \rho_{\bm{m}}^{\otimes Q} \mathcal{T}^{\otimes k}_{n,A}] \rangle, \\
    Z_0(\bm{m}) = \langle \text{Tr}[\check \rho_{\bm{m}}^{\otimes Q} ]\rangle.
\end{cases}
\end{equation}
Here, we have defined $Q \equiv k n +1$ and we left implicit that $\mathcal{T}_{n,A}^{\otimes k}$ acts as the identity permutation on the $Q$-th replica.

As we show in the following, in the limit of large local Hilbert space dimension it is possible to exactly compute the replica limit $k\to 0$ ($Q\to 1$) in Eq.~\eqref{eq:snZ} and formulate the entanglement entropy in terms of partition functions of a $Q=1$ replica of a classical two-dimensional statistical mechanics model.

\subsection{Specific model}
\begin{figure*}[t]
\includegraphics[width=0.65\textwidth]{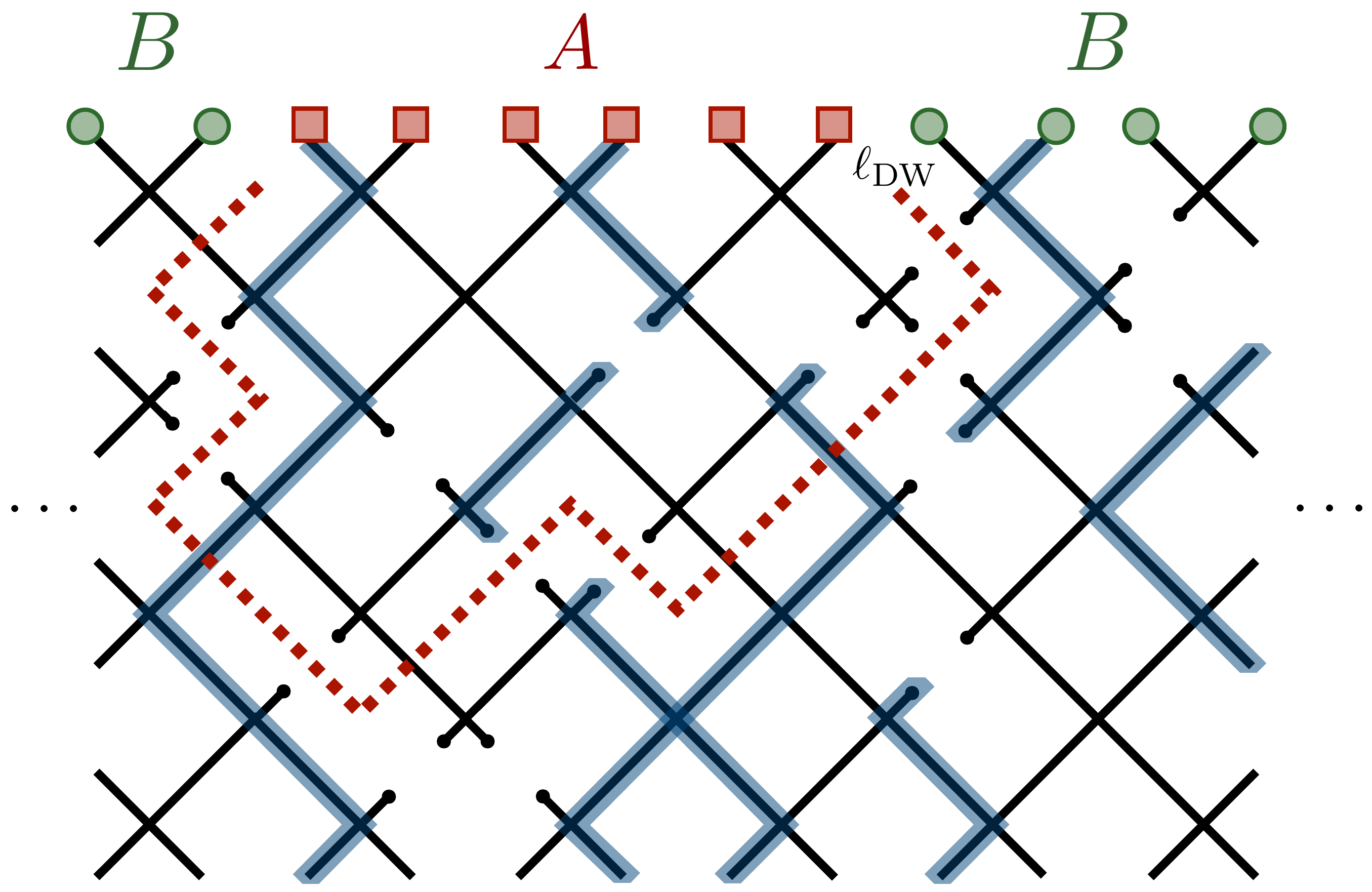}
\centering
\caption{Two dimensional statistical model describing the average Rényi-$n$ entropy of the monitored circuit of Sec.~\ref{sec:model} in the limit of large local Hilbert space dimension. The lattice vertices correspond to the averaged replicated unitary gates and bonds represent the sites of the quantum system. Broken bonds correspond to local projective measurements. On each vertex is defined a permutation degree of freedom $\sigma \in S_n$. The bipartition into subsystems $A\cup B$ imposes specific boundary conditions on the top layer, indicated by the red squares and green circles.  Red squares represent the cyclic permutation, while green circles denote the identity permutation. The vertices are constrained to be in the same permutation configuration as the boundary, creating a domain of length $\ell_{\rm DW}$ that cuts through the smallest number of unmeasured bonds (dashed red line). In the one replica limit, each bond hosts a charge variable $\alpha = \pm 1$  associated with the $U(1)$ symmetry of the circuit gates. Blue lines indicate the bonds with charge $\alpha=+1$. The dynamics of the charge follows a SSEP. 
When measurements explicitly break the symmetry, they act as local sources and sinks for the conserved charge, injecting or removing charge with equal probability, independently of the measurement outcome.}
\label{fig:SMmodel} 
\end{figure*}

Using the previous general framework, we now analyze the model described in Sec.~\ref{sec:model} in the limit of $d \to \infty$. This limit has been considered in many settings~\cite{nvh-18, zn-19, krps-18, rpk-18, mcculloch23, bca-20, jian20}, and in our case makes the average entanglement entropy analytically tractable via the two-dimensional statistical mechanics model illustrated in Fig.~\ref{fig:SMmodel}. In doing so, we follow the approach of Ref.~\cite{vasseur22}.

A convenient way to study the replicated random circuit in Eq.~\eqref{eq:Zs} is to vectorize the operators in a doubled Hilbert space using the Choi-Jamio\l{}kowski isomorphism~\cite{jamio72,choi75}. 
In particular, the unnormalized density matrix $\check{\rho}_{\bm{m}}$, evolving as $\check{\rho}_{\bm{m}} = K_{\bm{m}} \ket{\psi_0}\bra{\psi_0} K_{\bm{m}}^{\dagger}$, is mapped to a state $\ket{\check\rho_{\bm{m}}}$ in the doubled Hilbert space $\mathcal{H} \otimes \mathcal{H}^*$, with the corresponding evolution $\ket{\check\rho_{\bm{m}}} = (K_{\bm{m}}\otimes K_{\bm{m}}^*)\ket{\psi_0}^{\otimes 2}$.
In this approach, the partition functions in Eq.~\eqref{eq:Zs} can be written as the overlaps 
\begin{eqnarray}\label{eq:Z_ovelaps}
 Z_A(\bm{m})=\bra{\bcirc \bsquare} \langle K_{\bm{m}}^{\otimes Q}\otimes K_{\bm{m}}^{*\otimes Q}\rangle\ket{\psi_0}^{\otimes 2Q},\\
Z_0(\bm{m})=\bra{\bcirc \bcirc} \langle K_{\bm{m}}^{\otimes Q}\otimes K_{\bm{m}}^{*\otimes Q}\rangle\ket{\psi_0}^{\otimes 2Q},
\end{eqnarray}
where $\bra{\bcirc \bcirc}$ and $\bra{\bcirc \bsquare} $ are boundary states acting on the replicated doubled Hilbert space that encode the contractions at the final time layer. The former implements the trace independently in each replica, while the latter additionally inserts a cyclic permutation of the replicas belonging to subsystem $A$, according to the action of $\mathcal{T}^{\otimes k}_{n,A}$.

Since the random unitary gates are drawn independently, we can average each $Q$-fold replicated two-site gate contained in $K_{\bm{m}}$ separately. We denote this average as $W_{j,j+1} \equiv \langle U_{j, j+1}^{\otimes Q} \otimes U_{j, j+1}^{*\otimes Q} \rangle$, where we have explicitly written the site indices $j$ and $j+1$ on which the gate acts. This average can be evaluated by applying standard Weingarten calculus~\cite{weingarten-78, bb-96, cs-06} for the random unitaries $U_q$ in Eq.~\eqref{eq:gatedecomp}. In particular, in the limit $d\to\infty$, the averaged replicated gate $W_{j, j+1}$ becomes the projector~\cite{vasseur22}
\begin{equation}
\label{eq:BWvert}
    W_{j,j+1}  = \sum_{\sigma \in S_Q} \,\left(\prod_{k=1}^Q \frac{1}{d^2} \frac{\delta_{\alpha_{k},\beta_{k}} \delta_{\gamma_k,\delta_k}+\delta_{\alpha_k,\gamma_k} \delta_{\beta_k,\delta_k}}{2} \right)\ket{\bm{\gamma};\sigma}_j\ket{\bm{\delta};\sigma}_{j+1}\bra{\bm{\alpha};\sigma}_j \bra{\bm{\beta};\sigma}_{j+1}.
\end{equation}
The states $\ket{\bm{\alpha},\sigma}_j \in (\mathcal{H}_1 \otimes \mathcal{H}_1^*)^{\otimes Q}$ are defined on each site $j$ as 
\begin{equation}
\label{eq:permstates}
    \left\{\ket{\bm{\alpha};\sigma}_j \equiv \sum_{g_1,\cdots,g_Q=1}^d \ket{g_1\cdots g_Q \alpha_1\cdots \alpha_Q ,g_{\sigma(1)} \cdots g_{\sigma(Q)}    \alpha_{\sigma(1)} \cdots \alpha_{\sigma(Q)} }_j\right\},
\end{equation}
where $\sigma \in S_Q$ is a permutation among $Q$ elements, and $\alpha_k = \pm 1$ labels the charge sector in the $k$-th replica of site $j$. 
The states $\ket{g_k}$, $g_k=1, \dots, d$, form an 
orthonormal basis of the $\mathbb{C}^d$ sector in the 
$k$-th replica, while $\ket{\alpha_k}$ are the eigenstates of $Z_j$ acting on the corresponding $\mathbb{C}^2$ sector.
The numerator in Eq.~\eqref{eq:BWvert} enforces qubit-charge conservation for each replica pair $\{k,\sigma(k)\}_{k=1}^Q$, while the denominator corresponds to the diagonal Weingarten coefficient, i.e. the inverse of the dimension of the replicated charge sector in the limit $d\to\infty$.

In the brickwork circuit geometry, each site is acted upon by a gate in the preceding time layer and by another gate in the following one, as shown in Fig.~\ref{fig:sketch}. Consequently, in the absence of measurements, the outgoing state $\ket{\bm{\gamma};\sigma}_j$ of each averaged gate~\eqref{eq:BWvert} is contracted with the incoming state $ \bra{\bm{\alpha};\tau}_j$ of the adjacent averaged gate in the next time layer. The scalar product between the states~\eqref{eq:permstates} corresponding to different permutations reads
\begin{equation}
\label{eq:contr_perm}
    B_{\bm{\alpha} \bm{\beta}}(\sigma^{-1}\tau)\equiv\langle \bm{\alpha};\sigma|\bm{\beta};\tau\rangle = d^{Q - \ell(\sigma^{-1}\tau)}\prod_{j = 1}^{Q - \ell(\sigma^{-1}\tau)} \prod_{l \in C_j} \delta_{\alpha_l,\beta_l},
\end{equation}
where $\{C_j\}$ denote the disjoint cycles of the permutation $\sigma^{-1}\tau$ and $\ell(\sigma)$ is the Cayley distance of $\sigma$ from the identity permutation, i.e. the minimal number of transpositions required to generate $\sigma$. In the limit $d\to \infty$, the prefactor in Eq.~\eqref{eq:contr_perm} suppresses configurations with non-minimal Cayley distance $\ell(\sigma^{-1}\tau)$, so that the leading contribution corresponds to $\sigma=\tau$. The last term in Eq.~\eqref{eq:contr_perm} constrains the  incoming and outgoing charges $\alpha_k$ and $\beta_k$ to be equal within each cycle $C_j$ of $\sigma^{-1}\tau$.

The last component we need to consider in the circuit is the monitoring. Suppose that a given site has been measured, with possible outcomes $\ket{x,m}$, where $x=\pm 1$ denotes the eigenvectors of the Pauli matrix $X$ acting on the qubit and $m = 1,\cdots,d$ labels the measurement basis states for the qudit. As already noted, the precise choice of the qudit measurement basis is irrelevant, since the dynamics in the qudit sector is fully random and does not respect any symmetry. Let $\Pi_{x,m} = \ket{x,m} \bra{x,m}$ be the projector onto the outcome $\ket{x,m}$. When a site is measured, all its replicas are projected onto the same state via the replicated vectorized projector $\ket{\Pi_{x,m}}\bra{\Pi_{x,m}}$, with $\ket{\Pi_{x,m}} = (\ket{x,m} \otimes \ket{x,m})^{\otimes Q}$, which is inserted  between the two adjacent averaged gates. The effect of a measurement is then determined by the contraction
\begin{equation}
\label{eq:BWmeas}
    \langle \bm{\alpha};\sigma |\Pi_{x,m} \rangle \langle \Pi_{x,m} |\bm{\beta};\tau\rangle = \prod_{i = 1}^Q |\langle\alpha_i | x\rangle|^2  |\langle \beta_i | x\rangle|^2 = \frac{1}{4^Q}.
\end{equation}
Therefore, at the measured sites, different permutation degrees of freedom $\sigma$ and $\tau$ on the outgoing and incoming adjacent gates are not suppressed, in contrast to the unmeasured sites. Eq.~\eqref{eq:BWmeas} also shows that, due to the measurement, the incoming and outgoing charges $\bm{\alpha}$ and $\bm{\beta}$ on the site are completely unconstrained. As a result, the corresponding contraction is independent of the measurement outcome. This property considerably simplifies the computation of the average entanglement entropy, as we will see below. At the end of the section, we discuss how the situation changes for symmetry-breaking measurements different from $X$, for which Eq.~\eqref{eq:BWmeas} depends on the measurement outcome.

We now have all the necessary ingredients to define the classical statistical-mechanics model emerging in the large local Hilbert space dimension limit, from which we can compute the partition functions in Eq.~\eqref{eq:Z_ovelaps}, and hence the average entanglement entropy~\eqref{eq:snZ}.
The resulting model is defined on a two-dimensional tilted square lattice, as shown in Fig.~\ref{fig:SMmodel},  in which each vertex corresponds to a replicated gate with a degree of freedom $\sigma \in S_Q$. The unmeasured sites of the quantum system correspond to bonds of the lattice connecting neighboring vertices. Each bond carries $Q$ copies of the local qubit-charge variable $\alpha_k=\pm1$, with $k =1,\cdots,Q$. Whenever a measurement occurs, the corresponding bond is broken, so that the charges at its two ends become effectively unconstrained. This is the main difference compared to the case of measuring in the $Z$ basis, in which the charge at the broken bonds is fixed by the measurement outcome~\cite{vasseur22}. The Boltzmann weights of the statistical model on the vertices, on the bonds, and on the broken bonds are given by Eqs.~\eqref{eq:BWvert},~\eqref{eq:contr_perm}, and~\eqref{eq:BWmeas}, respectively. 

Eqs.~\eqref{eq:Z_ovelaps} correspond to the partition function of this statistical model with different boundary conditions imposed at the top layer. In the case of $Z_0(\bm{m})$ all the permutations at the top layer are fixed to the identity. Instead, for $Z_A(\bm{m})$, the permutations on subsystem $A$ are fixed to be cyclic.
In the limit $d\to\infty$, the Boltzmann weight~\eqref{eq:contr_perm} of an unmeasured bond forces the permutation degrees of freedom at the two vertices it connects to be identical. As a result, the different boundary conditions on $A$ and its complement create a domain wall separating a region of vertices in the cyclic permutation configuration from a region of vertices in the identity permutation configuration. This domain wall follows the path connecting the endpoints of $A$ on the top boundary that cuts the minimum number of unmeasured bonds~\cite{vasseur22}, corresponding to the red dashed line in Fig.~\ref{fig:SMmodel}. We will denote by $\ell_{{\rm DW}}(\bm{X})$ the number of unmeasured bonds that the domain wall cuts, which depends on the specfic realization of measument locations $\bm{X}$.

\subsection{Entanglement entropy from the statistical model}\label{subsec:ee_sm}

After summing over the measurement outcomes of the qudits and performing exactly the replica limit $k\to 0$ as shown in Ref.~\cite{vasseur22}, the trajectory-averaged Rényi entropy $\mathbb{E}_{\bm{m}} [S_n]$ for a given set of
measurements locations $\bm{X}$ can be written as a sum of a qudit contribution $S^{d}$ and a qubit one $S^{\rm{q}}$,
\begin{equation}
    \mathbb{E}_{\bm{m}} [S_n] = S^d + S^{\rm{q}},
\end{equation}
with
\begin{equation}
\label{eq:Sd}
    S^d =  \ell_{\rm{DW}} \log d,
\end{equation}
and
\begin{equation}
\label{eq:Sq}
    S^{\rm{q}} = \sum_{\bm{x}} Z_0(\bm{x}) \frac{1}{1-n} \log \left( \frac{Z^{(n)}_A(\bm{x})}{Z_0(\bm{x})^n}\right).
\end{equation}
The qudit entanglement contribution in Eq.~\eqref{eq:Sd} solely depends on the minimal cut length $\ell_{\rm{DW}}$, induced by the spatial bipartition into the subsystem $A$ and the complement. In terms of the statistical model, each measurements location $\bm{X}$ represents a percolation configuration. The length $\ell_{\rm{DW}}(\bm{X})$, when averaged over the realizations $\bm{X}$, displays a phase transition governed by the classical $2$D percolation universality class. This is the same universality class as the MIPT in monitored non-symmetric Haar circuits in the large-$d$ limit~\cite{srn19,bca-20, jian20}. The critical point is located at $p_c = 1/2$ and the correlation length critical exponent is $\nu = 4/3$ \cite{kesten,kesten1,grimmett}. For $p<p_c$, the averaged $\ell_{\rm DW}$ is extensive in the subsystem size while for $p>p_c$ is subsystem-size independent.

Let us now analyze the qubit contribution to entanglement. In Eq.~\eqref{eq:Sq}, $\bm{x}$ denotes the string of measurement outcomes for the qubit and $Z_0(\bm{x})$ corresponds to the Born probability of the outcome $\bm{x}$, which is given by the one replica version of the statistical model we have described. For this reason, we trivially have the normalization condition $\sum_{\bm{x}} Z_0(\bm{x}) = 1$. The partition function $Z_A^{(n)}(\bm{x})$ can be rewritten as a sum over measurement outcomes along the minimal-cut domain wall,
\begin{equation}
    Z_{A}^{(n)}(\bm{x}) = \sum_{\bm{\beta}} Z_A^{(1)}(\bm{x}; \bm{\beta})^n,
\end{equation}
where $Z_A^{(1)}(\bm{x};\bm{\beta})$ is the partition function of the same statistical model as $Z_0(\bm{x})$, with the qubit-charge configuration fixed on the unmeasured bonds along the minimal cut, $\bm{\beta} = \{\beta_i\}_{i = 1,\cdots,\ell_{\rm{DW}}}$. Therefore, the ratio
\begin{equation}\label{eq:p_beta}
    p_{\bm{\beta}} \equiv Z^{(1)}_A(\bm{x},\bm{\beta})/Z_0(\bm{x}),
\end{equation}
exactly corresponds to the joint probability distribution of the charges $\bm{\beta}$ along the minimal cut for the statistical model with partition function $Z_0(\bm{x})$.  

A major simplification occurs because of our choice of the measurement operator. At the level of the statistical model, a $X$ measurement acts as an impurity described by the Boltzmann weight~\eqref{eq:BWmeas} regardless of the measurement outcome. This means that the partition function $Z_0(\bm{x})$ is independent on the qubit measurement outcomes $\bm{x}$. Consequently, using the normalization condition $\sum_{\bm{x}} Z_0(\bm{x}) = 1$, we have $Z_0(\bm{x}) = 2^{-N_{\bm{X}}}$, where $N_{\bm{X}}$ is the total number of measurements for the configuration $\bm{X}$. 
The partition functions $Z_A^{(1)}(\bm{x};\bm{\beta})$ are also independent of $\bm{x}$.
The qubit contribution~\eqref{eq:Sq} to the average entanglement (with fixed measurement locations $\bm{X}$) then simplifies as
\begin{equation}
\label{eq:Sqsimpl}
    S^{\mathrm{q}} = \frac{1}{1-n} \log \left(\sum_{\bm{\beta}} p_{\bm{\beta}}^n \right) = \frac{1}{1-n} \log\left(\sum_{\bm{\beta}}\frac{Z_A(\bm{\beta})^n}{Z_0^n}\right).
\end{equation}
The determination of the average R\'enyi entropy of $p_{\bm{\beta}}$ over the measurement locations $\bm{X}$ still remains a difficult task, as each contribution to the average generally depends non-trivially on $\bm{X}$. To this end, we find it convenient to interpret the two-dimensional statistical model in terms of a one-dimensional Symmetric Simple Exclusion Process (SSEP), a well-known paradigmatic model for classical non-equilibrium stochastic processes~\cite{mallick-15}, but in the presence of defects disordered homogeneously in space and time.

\subsection{Statistical model as a disordered exclusion process}

The statistical model that emerges in the large-$d$ limit contains two types of degrees of freedom: permutation variables, associated with the replicated qudit sector, and binary charge variables, associated with the qubit sector. The  dynamics of the latter in the one-replica limit corresponds to a SSEP. The measurements in the $X$ basis act as defects located at the spacetime points $\bm{X}$, where the usual exclusion-process dynamics is locally modified.

We can represent the possible qubit-charge configurations by the set of occupation vectors $\ket{\bm{n}}$, with $n_j = 0,1$ for $j = 1,\cdots,L$. The probability distribution $p_{\bm{n}}(t)$ over configurations at time $t$ is encoded in the vector $\ket{P(t)} = \sum_{\bm{n}} p_{\bm{n}}(t) \ket{\bm{n}}$. This state has normalization condition $\braket{ \mathbbm{1} |P(t)} = \sum_{\bm{n}}p_{\bm{n}}= 1$, where we have introduced the unnormalized uniform state $\ket{\mathbbm{1}} = \sum_{\bm{n}} \ket{\bm{n}}$. In the absence of measurements, the state $\ket{P(t)}$ evolves according to the transfer matrix of the SSEP model. Owing to the brickwork structure of the circuit, the dynamics is implemented through alternating applications of even and odd transfer matrices,  i.e. $\ket{P(t)} = (\mathbb{T}_e \mathbb{T}_o)^t \ket{P(0)}$, where $\mathbb{T}_{e/o} = \prod_{j\in {\rm even}/{\rm odd}} T_{j,j+1}$. The two-site transfer matrix $T_{j, j+1}$ is obtained from the charge conservation term of the one-replica limit of the average random unitary gate~\eqref{eq:BWvert}, from which the effective SSEP description of the qubit sector emerges. It acts as
\begin{equation}
    \begin{cases}
        T\ket{00} = \ket{00},\\
        T \ket{10} = (\ket{10} + \ket{01})/2,\\
        T \ket{01} = (\ket{10} + \ket{01})/2, \\
        T \ket{11} = \ket{11}.
    \end{cases}
\end{equation}
Without monitoring, this model is known to display diffusive hydrodynamic transport of the conserved charge, given by the total occupation number $\sum_j n_j$, with fluctuations described by the Macroscopic Fluctuation Theory~\cite{mft-15}.

In the presence of monitoring in the $X$ basis, a measurement at site $j$ is implemented by the single-site operator 
\begin{equation}
\label{eq:M}
\begin{cases}
    M \ket{0} = (\ket{0} + \ket{1})/4, \\
    M \ket{1} = (\ket{0} + \ket{1})/4,
\end{cases}
\end{equation}
regardless of the outcome. This expression follows from the contraction in Eq.~\eqref{eq:BWmeas} in the replica limit $Q\to1$.
Eq.~\eqref{eq:ssep_rules} shows that, independent of the value of the charge at the site, the measurement maps the local state to a uniform probability distribution over the two possible charge values. This differs from the case in which the measurement is performed in the $Z$ basis, which pins the occupancy to the measurement outcome at the measured site~\cite{vasseur22, gmv26}. 
Notice that the measurement breaks the normalization $\langle \mathbb{I}|P(t)\rangle=1$ and therefore violates the conservation of probability, since an outcome has been selected (although irrelevant for our type of measurement). 

The dynamics of the monitored system is now given by a disordered transfer matrix with the deterministic layers of $\mathbb{T}_{e/o}$ interspersed by the operator $M$ applied in random locations $\bm{X}$. If we denote the full transfer matrix for a given disorder realization $\bm{X}$ as $\mathbb{T}_{\bm{X}}$, then the partition function $Z_0$ can be computed as the contraction 
\begin{equation}
Z_0 = \bra{\mathbbm{1}} \mathbb{T}_{\bm{X}}\ket{P(0)}.
\end{equation}
An important property of this disordered dynamics is that it preserves the uniform probability state $\ket{\mathbbm{1}}$ up to an overall proportionality factor determined by the number of measurement locations $N_{\bm{X}}$. This factor arises because the measurements break probability conservation. Specifically, since $T_{j, j+1}\ket{\mathbbm{1}}=\ket{\mathbbm{1}}$ and 
$M_j\ket{\mathbbm{1}}=2^{-1}\ket{\mathbbm{1}}$, then
$T_{\bm{X}} \ket{\mathbbm{1}}  = 2^{-N_{\bm{X}}}\ket{\mathbbm{1}}$. 
This result implies that $Z_0 = 2^{-N_{\bm{X}}}$, independently of the initial distribution $\ket{P(0)}$, consistent with the normalization condition of the Born probability discussed in Sec.~\ref{subsec:ee_sm}. 

To compute the probability distribution $p_{\bm{\beta}}$ in Eq.~\eqref{eq:Sqsimpl},
we also need the partition function $Z_A(\bm{\beta})$, which is described by the same SSEP as $Z_0$, but with fixed occupancies $\bm{\beta}$ on the unmeasured bonds along the minimal cut associated with $A$. We can fix these occupancies in the SSEP transfer matrix by inserting a projector on each bond of the minimal cut that enforces the charges to be $\bm{\beta}$. If we denote by $\mathbb{T}^{\bm{\beta}}_{\bm{X}}$ the corresponding total transfer matrix, then
\begin{equation}\label{eq:p_beta_T}
    p_{\bm{\beta}} = 2^{N_{\bm{X}}} \bra{\mathbbm{1}} \mathbb{T}^{\bm{\beta}}_{\bm{X}} \ket{P(0)},
\end{equation}
where we have used $Z_0 = 2^{-N_{\bm{X}}}$.

To analyze Eq.~\eqref{eq:p_beta_T}, we fix, for simplicity, the initial state $\ket{P(0)}$. A convenient choice is the uniform distribution over the occupations, i.e. $\ket{P(0)} = 2^{-L} \ket{\mathbbm{1}}$. This corresponds to initialize the qubit sector of the quantum system in the state $2^{-L/2} \bigotimes_{j=1}^L (\ket{0}_j + \ket{1}_j)$. 
Because of the form of the measurements~\eqref{eq:M}, the state $\ket{\mathbbm{1}}$ is invariant under them and all single-site charge probabilities are uniform, $p_{\beta_j} = 1/2$. Hence, 
in the transfer matrix $\mathbb{T}_{\bm{X}}^{\bm{\beta}}$, we insert the local projectors
\begin{equation}
P_{\beta_j}=\frac{1+\beta_j \,Z_j}{2}
\end{equation}
on the unmeasured bonds belonging to the minimal cut. If, for a specific realization $\bm{X}$, all the minimal cut bonds are causally disconnected by the $X$-monitored SSEP dynamics, the corresponding contributions factorize, and the probability reduces to
\begin{equation}
    p^*_{\bm{\beta}} = 2^{-\ell_{\rm{DW}}} \prod^{\ell_{\rm{DW}}}_{j=1}\bra{\mathbbm{1}} P_{\beta_j}\ket{\mathbbm{1}}_j = 2^{-\ell_{\rm{DW}}}.
\end{equation}
Inserting this result into Eq.~\eqref{eq:Sqsimpl}, we obtain the maximal entropy, $S^{\rm{q}} = \ell_{\mathrm{DW}}\log 2$, for the qubit sector. As for the qudit contribution~\eqref{eq:Sd}, the minimal cut length $\ell_{\rm{DW}}$ is the only parameter controlling the entanglement. This situation, in which the bonds with fixed charge $\bm{\beta}$ are causally disconnected, occurs, for example, when the minimal cut is completely horizontal, as expected for small values of the measurement rate $p\approx 0$.

In practice, in a generic measurement configuration $\bm{X}$, charge correlations may survive along the minimal cut. However, we will now show that symmetry-breaking measurements suppress such correlations, resulting in a finite typical correlation length $\xi_{\text{typ}} < \infty$ for any measurement rate $p$. As a consequence, the previous approximate picture of uncorrelated charges along the minimal cut remains a good approximation. Let us consider the connected two-point function of the local occupation number $C_{\bm{X}}(z,t) \equiv \langle n(z,t)\, n(0,0)\rangle_c\geq 0 $, where $z$ denotes the spatial coordinate, evaluated in the uniform state $2^{-L} \ket{\mathbbm{1}}$ for a given measurement configuration $\bm{X}$. In App.~\ref{sec:app1}, we obtain that the average correlation is exponentially suppressed in time,
\begin{equation}
\label{eq:avgC}
    \mathbb{E}_{\bm{X}}[ C_{\bm{X}}(z,t) ] \leq \frac{1}{4} (1-p)^{2t} = \frac{1}{4} e^{-t/\xi_p}, \qquad \text{if } |z|\leq t
\end{equation}
and $0$ otherwise. Here, we have introduced~$\xi^{-1}_p = -2\log(1-p) > 0$ as the average correlation length, which is finite for any the measurement rate $p>0$. This result essentially tells us that charge correlations between two space-time points only survive if the SSEP charge histories contributing to $C_{\bm{X}}(z,t)$ encounter no measurements. When treating disordered systems, it is important to distinguish the average correlation~\eqref{eq:avgC} from the typical one~\cite{fisher94,fisher95,fy98,im05,rm09}, which is defined as $C_{\text{typ}}(z,t) \equiv \exp(\mathbb{E}_{\bm{X}}[\log C_{\bm{X}}(z,t)])$, where we are conditioning on the non-zero values of $C_{\bm{X}}(z,t)$. In our case, using Eq.~\eqref{eq:avgC} and Jensen's inequality, we have
\begin{equation}
    -\frac{1}{t} \mathbb{E}_{\bm{X}}[\log C_{\bm{X}}(z,t)] \geq \xi_p^{-1},
\end{equation}
which shows that the typical correlation length is necessarily finite as well.
Both the average and typical charge correlation lengths are constrained to remain finite for any $p>0$ as a consequence of the symmetry-breaking measurement. See also App.~\ref{app:sm} for further analysis on this point; in particular, we prove that, with probability one over $\bm{X}$, the correlation $C_{\bm{X}}(z,t)$ asymptotically decays exponentially in time, with an inverse correlation length bounded from below by $\xi_p^{-1}$.

As a result, for all $p>0$, the only possible diverging length scale in the statistical model is the percolation length scale. The finiteness of the typical charge correlation length implies that the charge distribution $p_{\bm{\beta}}$ can be coarse-grained into $O( \ell_{\mathrm{DW}}/\xi_{\text{typ}})$ effectively independent variables along the minimal cut. Consequently, in terms of the expression for $S^{\mathrm{q}}$ in Eq.~\eqref{eq:Sqsimpl}, the scaling behavior of $S^{\mathrm{q}}$ is controlled solely by the minimal-cut length $\ell_{\mathrm{DW}}$, as in the qudit contribution~\eqref{eq:Sd}. 
Based on this argument, both the qubit and qudit sectors undergo an entanglement phase transition in the same universality class as the MIPT in monitored non-symmetric Haar circuits in the large-$d$ limit. This is a direct consequence of symmetry-breaking measurements inducing a finite correlation length in the charge sector.
By contrast, when measurements preserve the $U(1)$ symmetry, a charge sharpening transition occurs at a measurement rate $p_\#<p_c$~\cite{vasseur22, barratt22}. In that case, the average charge correlation length is finite for $p>p_\#$, and the average correlation $\mathbb{E}_{\bm{X}}[C_{\bm{X}}(z, t)]$ decays exponentially, while it exhibits algebraic decay for $p<p_\#$.

Let us comment on the specific choice of the symmetry-breaking measurement. As described above, the choice of measuring $X$ leads, in the limit $d \to \infty$, to a uniform Born probability over trajectories, thus greatly simplifying the discussion. However, as we discuss in App.~\ref{sec:app2}, and following a similar argument to that presented in this section, we expect that any symmetry-breaking measurement, e.g. in the basis of eigenstates of $Z \cos \theta + X \sin \theta$ with $0<\theta < \pi$, generates a finite typical charge correlation length $\xi_p^{-1}(\theta) = -2 \log(1-p\sin^2 \theta)$ for any $p>0$. This implies that, for any small amount of symmetry-breaking in the measurement basis, the critical behavior of the MIPT belongs to the universality class of non-symmetric Haar random circuits. For very small values of $\theta$, the typical charge correlation length becomes parametrically large, which may lead to crossover phenomena in finite-size systems.

Finally, our result also applies to a different measurement protocol in which every local measurement is performed in an independent random basis of the full local Hilbert space $\mathbb{C}^2\otimes \mathbb{C}^d$. As we show in App.~\ref{app:rndbas}, in the large-$d$ limit, this protocol exactly yields the same Boltzmann weight in Eq.~\eqref{eq:BWmeas} as measurements performed in the $X$ eigenbasis. The random measurement basis can be understood as a way of modeling noise in the monitored system. Our results complements other studies in which the role of noise in breaking symmetries at measurement-induced phase transitions has also been widely explored~\cite{bca-21, iom23,dias23,wba22, liu24,liu24-2}.

In the following, we analyze the same scenario in two different models of monitored quantum circuits at finite-Hilbert-space dimension, where no analytical treatment is available. The percolation universality is known to apply strictly only in the limit $d \to \infty$\cite{jian20, zabalo22}. Nevertheless, we observe that the intuition obtained from the large-$d$ limit regarding the role of symmetry-breaking measurements remains valid in finite-dimensional monitored systems, where a complete characterization of the MIPT universality class is still lacking.

\section{Finite-size scaling for a monitored qubit chain}
\label{sec:qubitnum}

In the previous section, we showed that, in the limit of large local Hilbert-space dimension $d\to\infty$, the measurement-induced phase transition of a $U(1)$-symmetric Haar-random circuit subject to symmetry-breaking measurements belongs to the two-dimensional percolation universality class, just as in monitored Haar-random circuits without conservation laws. In this section, we study the same model, defined in Sec.~\ref{sec:model}, but for $d = 1$, which corresponds to a one-dimensional chain of qubits with local Hilbert space $\mathcal{H}_1 = \mathbb{C}^2$. 
In this case, the effective statistical-mechanics description is analytically intractable. We therefore resort to a finite-size scaling analysis of the monitored circuit. We numerically simulate the exact dynamics of the system and sample many quantum trajectories. By doing so, we are able to reach up to $L = 24$ sites in the qubit chain.

Following Refs.~\cite{gh20, zabalo20, vasseur22}, to estimate the critical properties of the entanglement transition, we compute the tripartite mutual information, also known as topological entanglement entropy~\cite{kp06,lw06}, defined as 
\begin{equation}\label{eq:I_3}
\begin{split}
    I_{3,n}(A,B,C) &= S_n(A) + S_n(B) + S_n(C)\\& -S_n(A \cup B) - S_n(A \cup C) -S_n(B \cup C) + S_n(A \cup B \cup C),
\end{split}
\end{equation}
where, for a system of $L$ qubits, $A,B,C$ are three contiguous regions each with $L/4$ qubits. For the numerical estimation of critical properties, the tripartite mutual information is more convenient than the entanglement entropy of a single subsystem, as it cancels, by construction, the expected logarithmic dependence on the system size $L$ at the critical point, which is known to induce systematic errors in finite-size scaling analyses~\cite{gh20, zabalo20}. At the same time, $I_{3,n}$ still exhibits a transition from a volume-law phase to an area-law phase across the MIPT, remains finite at the critical point, and is therefore sensitive to the transition.
For measurement rates $p$ close to the critical one $p_c$, we expect the average over quantum trajectories (including measurement locations) $\mathbb{E}[I_{3,n}]$ to satisfy a finite-size scaling behavior. As generally observed for MIPTs, we assume the dynamical exponent $z=1$ and thus fix the time proportional to the system size, $t\propto L$. The expected scaling behavior at a MIPT is then given by
\begin{equation}
\label{eq:fssI3}
    \mathbb{E}[I_{3,n}] = \mathcal{I}_n\left[(p-p_c)L^{1/\nu}\right] + O(L^{-\omega}),
\end{equation}
where $\nu$ is the correlation length critical exponent and $\omega$ the exponent associated with the largest irrelevant scaling field. The function $\mathcal{I}_n(x)$ is a universal scaling function, meaning that it depends only on the universality class of the transition, up to non-universal rescalings of its argument~\cite{fisher98,cardy}. As a consequence, the value $\mathcal{I}_n(0)$ is universal and can serve as a useful diagnostic to distinguish different universality classes.

\begin{figure*}[t]
\includegraphics[width=0.7\textwidth]{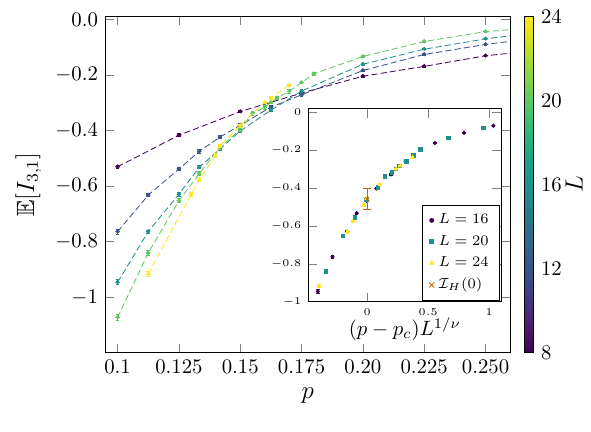}
\centering
\caption{Trajectory-averaged tripartite mutual information $\mathbb{E}[I_{3,1}]$, for $n=1$, in the stationary state at $t\sim 2L$ of the $U(1)$-symmetric Haar random circuit with measurements in the $X$ basis, as a function of the measurement rate $p$ and for increasing system size $L$. We average over $\sim 2\times 10^3$ quantum trajectories. The errorbars correspond to one standard deviation of the mean. For $L \geq 16$, the data develop a clear crossing as the system size increases, signaling the presence of an MIPT at that measurement rate. Inset: Finite-size scaling of $\mathbb{E}[I_{3,1}]$ according to the scaling ansatz in Eq.~\eqref{eq:fssI3}, using the optimal $p_c$ and $\nu$ from Eq.~\eqref{eq:num_pc_nu}, obtained as described in App.~\ref{app:num}. For $L \geq 16$, we observe that the data collapse onto the universal scaling function $\mathcal{I}_n(x)$. The red cross corresponds to the estimate of the universal value $\mathcal{I}^H_{n=1}(0)$ for a generic Haar-random circuit obtained in Ref.~\cite{zabalo20}.}
\label{fig:I3n1} 
\end{figure*}

We numerically study the monitored circuit defined in Sec.~\ref{sec:model} for $d=1$ by exactly evolving the initial state $\ket{\psi_0} =2^{-L/2} \bigotimes_{j=1}^L (\ket{0}_j + \ket{1}_j)$. For each gate realization and quantum trajectory, we compute the tripartite mutual information $I_{3,n}$ at time $t \sim 2 L$ (in units of full even/odd circuit layers), ensuring that the steady state has been reached. In Fig.~\ref{fig:I3n1} we show the results we obtain for R\'enyi index $n=1$, i.e. considering in Eq.~\eqref{eq:I_3} the von Neumann entropy, as a function of the measurement rate $p$ for different system sizes $L$.
In the area-law phase, $p>p_c$, the entanglement entropy saturates to a constant for large $L$, proportional to the number of subsystem endpoints; therefore, we expect the tripartite mutual information to vanish in the thermodynamic limit. Conversely, in the volume-law phase, $p<p_c$, since subsystems $A$, $B$ and $C$ have the same size, the tripartite mutual information should be negative and scale linearly with $L$. The results in Fig.~\ref{fig:I3n1} exhibit two different regimes that are consistent with these behaviors.

According to the scaling hypothesis~\eqref{eq:fssI3}, and neglecting the subleading corrections due to irrelevant RG perturbations, the critical measurement rate corresponds to the crossing point of the data for different system sizes. In Fig.~\ref{fig:I3n1}, we observe a clear crossing that shows a small drift with increasing $L$, due to finite-size effects.
Determining the optimal collapse according to Eq.~\eqref{eq:fssI3} (see Appendix~\ref{app:num} for details of the numerical procedure), we obtain the following estimates for the critical measurement rate and the correlation-length exponent:
\begin{equation}\label{eq:num_pc_nu}
   p_c = 0.143(3), \quad \nu = 1.3(2).
\end{equation}
The critical measurement rate $p_c$ is a non-universal quantity that depends on the specific model under consideration. 
We obtain a smaller  value of $p_c$ than in the MIPT of a generic Haar-random circuit because $U(1)$-symmetric gates generate, on average, less entanglement.
The obtained $\nu$ is compatible with both the best current estimates for the Haar random circuit, $\nu_H = 1.2(2)$~\cite{zabalo20}, and for the $U(1)$-symmetric Haar circuit with symmetric local measurements, $\nu_{U(1)} = 1.32(6)$~\cite{vasseur22}, which were determined in previous studies using the tripartite mutual information. In all these settings, $\nu$ suffers from large relative errors, which prevent a clear distinction between universality classes and the possible identification of new ones. However, in our model we obtain the universal value $\mathbb{E}[I_{3,1}(p_c)] \simeq -0.45(5)$, compatible with the estimate of Ref.~\cite{zabalo20} for Haar random circuits. In the inset of Fig.~\ref{fig:I3n1}, we show the data collapse of the results in the main panel for different system sizes when plotted as a function of $(p-p_c)L^{1/\nu}$, using the values of $p_c$ and $\nu$ reported in Eq.~\eqref{eq:num_pc_nu}.

\begin{figure*}[t]
\includegraphics[width=0.9\textwidth]{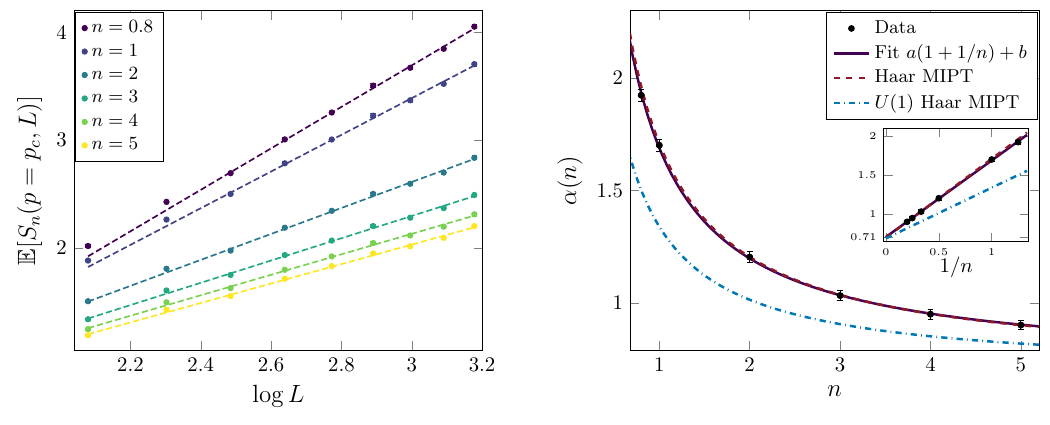}
\centering
\caption{Half-system R\'enyi-$n$ entanglement entropy $\mathbb{E}[S_n]$ averaged over $\sim 10^3$ quantum trajectories at the estimated critical point $p_c = 0.143$ in the $U(1)$-symmetric Haar random circuit with symmetry-breaking measurements in the $X$ basis. Left panel: Logarithmic growth of $\mathbb{E}[S_n(p_c)]$ as a function of the system size $L$ for various R\'enyi indices. The symbols are the exact numerical results, while dashed lines represent the fits of the function $\alpha(n) \log L + \beta(n)$ to the data. The errorbars are of the order of the symbol size and correspond to one standard deviation of the mean. Right panel: Symbols represent the values of the coefficient $\alpha(n)$ obtained from the fits as a function of the R\'enyi index $n$. The violet solid curve corresponds to the fit of the CFT-like function $\alpha(n) = a(1+1/n) + b$ to the symbols. The red dashed curve represents the same function, using for $a$ and $b$ the estimates reported in Ref.~\cite{zabalo20} (see also main text) for the monitored Haar-random circuit. The blue dash-dotted curve is the same function with the values of $a$ and $b$ estimated for the $U(1)$-symmetric Haar circuit with symmetric measurements in Ref.~\cite{vasseur22} (see also main text). Inset: Same data shown in the right panel but as a function of $1/n$. The results for the $U(1)$ Haar random circuit subject to symmetry-breaking measurements indicate that its critical point is compatible with the generic Haar MIPT universality class.}
\label{fig:Sncrit} 
\end{figure*}

A more stringent test of the universality class is the R\'enyi-index dependence of the entanglement entropy at the critical point. At $p = p_c$, the average half-system entanglement entropy shows a logarithmic growth in the system size $\mathbb{E}[S_n(p_c)] = \alpha(n) \log L$. The factor $\alpha(n)$ and its Rényi-index dependence are universal and more sensitive to the universality class~\cite{vasseur22}. In fact, both for the non-symmetric Haar random circuit and the symmetric one with symmetric measurements, $\alpha(n)$
admits the functional form
\begin{equation}\label{eq:alpha_n}
\alpha(n)=a(1+1/n)+b.
\end{equation}
For not symmetric Haar random circuits, the fit of the exact numerical data to this function gives $a_H=1.01$, $b_H=-0.31$~\cite{zabalo20}, while for $U(1)$ Haar circuits with symmetric monitoring, $a_{U(1)}=0.65(1)$ and $b_{U(1)}=0.04(1)$~\cite{vasseur22}. Up to the offset $b$, Eq.~\eqref{eq:alpha_n} has the same R\'enyi-index dependence as the entanglement entropy of a CFT ground state~\cite{cc-04}. In Fig.~\ref{fig:Sncrit} we show the numerical results obtained in our model for the  average R\'enyi-$n$ entanglement entropy for $n\leq 5$ at the critical point $p_c \simeq 0.143$ estimated above. In the left panel, we plot them as a function of $\log L$. The dashed lines are the fit of the function $\alpha(n)\log L+\beta(n)$. In the right panel, we plot the coefficients $\alpha(n)$ obtained from the fits as a function of $n$ (symbols). The solid curves correspond to fitting Eq.~\eqref{eq:alpha_n}, yielding $a=0.98(3)$ and $b=-0.27(4)$. The dashed and dot-dashed curves show Eq.~\eqref{eq:alpha_n} evaluated using the values of $a$ and $b$ reported previously for the Haar random circuit and the $U(1)$ Haar  circuit with symmetric measurements, respectively. We observe a striking agreement with the universality class of the non-symmetric Haar random circuit.

This result provides the main supporting evidence that, also at finite local Hilbert-space dimension, symmetry-breaking projective measurements amount to a relevant perturbation of the symmetric MIPT, which at large scales flows to the entanglement transition of the corresponding circuit without symmetry. One way to confirm this hypothesis is by verifying that the same universal results are obtained when measurements are performed in a rotated basis. However, our numerical analysis of the monitored Haar qubit circuit is limited by the relatively small system sizes accessible through exact simulation. Based on the large-$d$ analysis of the previous section, we therefore expect that introducing a tilted symmetry-breaking measurement would increase finite-size effects, requiring larger system sizes to reliably resolve the critical point.
Indeed, in the $d\to\infty$ limit, we found that the qubit sector is characterized by a finite correlation length $\xi(\theta)$, which increases as the measurement basis approaches the symmetry-preserving $Z$ basis ($\theta\to0$). The resulting growth of $\xi(\theta)$ enhances finite-size effects and delays the onset of the asymptotic scaling regime.

Nonetheless, although the results are compatible with the intuition developed in Sec.~\ref{sec:statmech} in the large-$d$ limit, we cannot fully exclude systematic finite-size effects in the estimates of the critical parameters. In the next section, we study the same setting in a monitored symmetric stabilizer circuit, which provides better control over finite-size effects and also allows us to test our hypothesis in an independent model.

\section{Finite-size scaling for a monitored stabilizer circuit}
\label{sec:stab}

The main conclusion of the previous sections is that, when monitoring a $U(1)$-symmetric circuit with symmetry-breaking measurements, the universality properties of the MIPT coincide with those of the non-symmetric Haar random circuit. In this section, we test whether this remains true in a different circuit model. 

A fundamental result in quantum many-body physics is the Gottesman-Knill theorem~\cite{gottesman97,gottesman98,gottesman98-2}. A consequence of this property is that quantum dynamics generated by Clifford gates (i.e., elements of the Clifford group, generated by the Hadamard, S, and CNOT gates) can be efficiently simulated classically by tracking the evolution of the Pauli strings that stabilize the quantum state~\cite{dm03,ag04,ab06}. The entanglement entropy of a generic region $A$ of the system can also be computed with polynomial complexity from the knowledge of the stabilizers~\cite{fattal04,hiz05,hiz05-2}.
This result allows us to access large size systems (in our simulations we reach up to $L=2048$ qubits) without the need to consider the large local Hilbert space dimension limit. 

In our case, we consider brickwork stabilizer circuits, as the one in Fig.~\ref{fig:sketch}, in which the two-qubit unitary gates are independently drawn with uniform probability from the subgroup of the Clifford group that preserves the magnetization $Z_j + Z_{j+1}$. This subgroup has been recently studied in the literature~\cite{rlp23,my23,marvian24,mitsuhashi25}. It consists of only $64$ possible gates out of all the $11520$ two-qubit Clifford group (modulo an irrelevant phase). We can parametrize a generic gate $U_{jk}$ of the $U(1)$-symmetric Clifford subgroup acting on two qubits $j$ and $k$ as \cite{my23}
\begin{equation}
\label{eq:u1Cliff}
    U_{jk} = \textrm{CZ}^{\mu} S^a_j S^b_k \,\textrm{SWAP}^\nu,
\end{equation}
where $\textrm{CZ}$ is the controlled Z gate, $S_j$ is the phase gate acting on the $j$-th qubit, and $\textrm{SWAP}$ is the swap gate. The exponents in Eq.~\eqref{eq:u1Cliff} can assume the values $\mu,\nu = 0,1$ and $a,b = 0,\cdots,3$. As in the Haar-random circuits considered in the previous sections, after each layer of unitary gates, each qubit is projectively measured in the $X$ basis with probability $p$, thereby explicitly breaking the conservation law of the unitary dynamics. This choice of measurement basis preserves the system in a stabilizer state at all times, rendering the computation of the entanglement entropy for each quantum trajectory classically tractable with polynomial complexity.

As proven in Ref.~\cite{ha24}, if the measurements are performed in the $Z$ basis, the stationary state lies in an area-law entangled phase for any finite $p>0$, and no MIPT occurs. We now show that symmetry-breaking measurements stabilize the volume-law phase at small finite values of the measurement rate $p$, leading to a MIPT in the system. According to the hypothesis put forward in this paper, we expect the universality class of the entanglement transition to coincide with that of monitored stabilizer circuits with gates drawn from the full Clifford group, which generically break the symmetry. The latter transition has already been widely studied in the literature~\cite{gh20,zabalo20,sierant22}, and we can verify the compatibility of the critical parameters between the two circuits.

\begin{figure*}[t]
\includegraphics[width=0.96\textwidth]{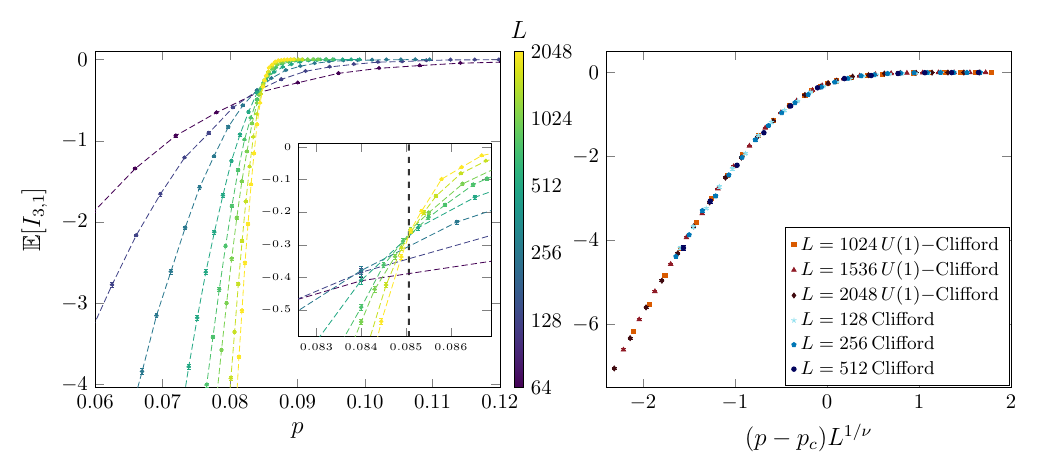}
\centering
\caption{Left panel: Tripartite mutual information $I_{3,1}$, for $n=1$, averaged over $\sim 6\times 10^3$ quantum trajectories, in the stationary state at $t\sim 2L$, 
of the symmetric random stabilizer circuit with projective measurements in the $X$ basis, as a function of the measurement rate $p$ and different system sizes $L$. The errorbars are of the order of the symbol size and correspond to one standard deviation of the mean. Inset: Zoom of the data around the critical point. We observe a drift of the crossing, signaling the MIPT, toward larger values of $p$ up to $L\sim 512$. For this reason, we perform the finite-size scaling using the data for $L\geq 768$. Right panel: Finite-size scaling of $\mathbb{E}[I_{3,1}]$ using the ansatz~\eqref{eq:fssI3} and the optimal $p_c$ and $\nu$ from Eq.~\eqref{eq:clifford_pc_nu}. As a comparison, we include data from a monitored non-symmetric stabilizer random circuit. In this case, we rescaled the scaling variable $(p-p_c)L^{1/\nu}$ by a non-universal factor $\simeq 0.94$. No vertical rescaling has been introduced as the tripartite mutual information is adimensional. The scaling functions of the two circuits agree for the values considered here, indicating that their critical points belong to the same universality class.
}
\label{fig:I3Cliff} 
\end{figure*}

We repeat the finite-size scaling analysis of the previous section, using the numerical methods described in App.~\ref{app:num}. All numerical simulations of the stabilizer circuit were performed using the Python library Stim~\cite{gidney21}. In the left panel of Fig.~\ref{fig:I3Cliff}, we show the results for the average tripartite mutual information~\eqref{eq:I_3}, taking $n=1$, as a function of the measurement rate $p$ and different system sizes $L$. We consider the subsystems $A$, $B$ and $C$ of equal size $L/4$. We observe two clear regimes: above certain finite $p$, $\mathbb{E}[I_{3,1}]$ tends to zero as $L$ increases, indicating area-law behavior, while below this value of $p$, it grows linearly with $L$, signaling a volume-law phase. The data crossing identifies the critical measurement rate $p_c$ of the MIPT. The inset of the left panel of Fig.~\ref{fig:I3Cliff} shows a zoom of the crossing region. While a finite-size drift of the crossing point is visible for small values of $L$, it converges as $L$ increases.  Applying the optimization procedure in App.~\ref{app:num} using the scaling ansatz~\eqref{eq:fssI3}, we obtain the estimates:
\begin{equation}\label{eq:clifford_pc_nu}
 p_c^C = 0.0851(1), \quad  \nu^C = 1.27(3).
\end{equation}
The critical measurement rate is smaller than that of the non-symmetric random Clifford circuit, $p_c^* = 0.15995(10)$~\cite{sierant22}, as expected from the lower entangling power of the symmetric gates. The correlation-length exponent is compatible with the one obtained in the non-symmetric random Clifford circuit, $\nu^*=1.260(15)$~\cite{sierant22}, supporting our hypothesis on the effect of symmetry-breaking measurements in a symmetric evolution. In the right panel of Fig.~\eqref{fig:I3Cliff}, we show the collapse of the data from the left panel as a function of $(p-p_c)L^{1/\nu}$, employing the values of $p_c$ and $\nu$ estimated in Eq.~\eqref{eq:clifford_pc_nu}. 
In the same figure, we also compare the collapsed data with those of the non-symmetric stabilizer circuit. The results for the non-symmetric circuit are plotted as a function of $a(p-p_c)L^{1/\nu}$, where $a$ is a non-universal rescaling factor. The data from the two circuits show a perfect collapse after this non-universal rescaling of the scaling variable, indicating that the corresponding critical points belong to the same universality class. This result is particularly remarkable, as the same critical theory is reproduced using the considerably smaller set of $U(1)$-symmetric Clifford gates.

\begin{figure*}[t]
\includegraphics[width=0.6\textwidth]{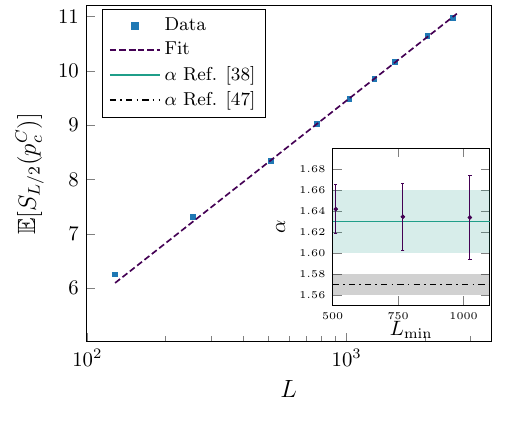}
\centering
\caption{Logarithmic growth of the half-system entanglement entropy $S_{L/2}$ averaged over $\sim 10^4$ quantum trajectories in the stationary state at $t\sim 2L$ of the symmetric random stabilizer circuit with projective measurements in the $X$ basis at the estimated critical point $p_c^C$ in Eq.~\eqref{eq:clifford_pc_nu}. Symbols correspond to the numerical data, while errorbars are smaller than the symbol size. The dashed line is the result of a linear fit to the function $\alpha \log L + \beta$. Inset: Comparison between our estimate of $\alpha$ and the corresponding values for the non-symmetric Clifford MIPT reported in Ref.~\cite{zabalo20} (green line) and Ref.~\cite{sierant22} (black dash-dotted line). In both cases, the shaded uncertainty band represents the one-standard-deviation confidence interval. The purple symbols represent our estimates as a function of the minimum system size $L_{\text{min}}$ included in the fit. Our results are compatible with both estimates for the non-symmetric case within two standard deviations.}
\label{fig:ceff} 
\end{figure*}

A well known property of stabilizer states is their flat entanglement spectrum, which implies that all R\'enyi entropies are equal~\cite{fattal04}. For this reason, at the critical point of the MIPT, the Rényi entanglement entropies can not display the ground state CFT-like behavior~\eqref{eq:alpha_n} found in the previous section for the Haar circuit. Nevertheless, at the critical point, the average half-system entanglement entropy remains logarithmic in the system size, $\mathbb{E}[S_n(p_c)] = \alpha \log L + b$, with the coefficient $\alpha$ expected to be universal~\cite{zabalo20,sierant22,li24}. In Fig.~\ref{fig:ceff}, we show the logarithmic growth of the half-system entanglement entropy and the estimation of $\alpha$ in the symmetric random stabilizer circuit with projective measurements in the $X$ basis. In the inset, we compare our prediction with the coefficient $\alpha$ of the non-symmetric Clifford MIPT obtained in Refs.~\cite{zabalo20,sierant22}, finding a good agreement. This result is consistent with the logarithmic coefficient being universal and provides further evidence that the MIPT in symmetric stabilizer circuits with symmetry-breaking projective measurements belongs to the same universality class as the non-symmetric stabilizer circuits.

Here, we restricted ourselves to a circuit with finite local Hilbert space dimension. In Ref.~\cite{li24}, a Weingarten formula analogous to Eq.~\eqref{eq:BWvert} was derived for the average of replicated Clifford gates. Using this result, a statistical model for monitored brickwork stabilizer circuits was obtained, which is analytically tractable in the large local Hilbert-space limit. In principle, such a model might be generalized to the case of symmetric Clifford gates and (symmetry-breaking) measurements.

\section{Conclusions}
\label{sec:Concl}

In this work, we studied measurement-induced phase transitions in monitored quantum circuits in the presence of symmetries and symmetry-breaking measurements. We considered $U(1)$-symmetric Haar and stabilizer random circuits, and investigated the effect of local projective measurements that explicitly break the conservation law of the unitary dynamics.

We showed that such symmetry-breaking measurements constitute a relevant perturbation at the measurement-induced critical point. As a result, the entanglement transition flows to the same universality class as that of the corresponding monitored circuit without symmetries. This is supported numerically at finite local Hilbert-space dimension, using both $U(1)$-symmetric Haar and stabilizer random circuits, and analytically in the large local Hilbert space dimension limit of the $U(1)$ Haar-random circuit. Remarkably, in monitored $U(1)$-symmetric stabilizer circuits, the MIPT is absent when the measurements preserve the symmetry, whereas symmetry-breaking measurements stabilize the volume-law phase at small but finite measurement rates, giving rise to an MIPT that belongs to the same universality class as that of non-symmetric stabilizer circuits.

In the large local Hilbert space dimension limit of the Haar-random circuit, the trajectory-averaged entanglement entropy can be mapped exactly, via the replica trick, onto a classical statistical-mechanics model, in which vertices correspond to the circuit gates and bonds to the sites of the quantum system. In this mapping, local projective measurements, regardless of whether they preserve or break the symmetry, are represented by broken bonds, while the bipartition of the system imposes boundary conditions that induce a domain wall whose minimal-cost path through the lattice crosses the smallest possible number of unmeasured bonds. The $U(1)$ symmetry of the gates introduces additional degrees of freedom associated with the transport of a conserved charge through the lattice, described by a symmetric simple exclusion process. Symmetry-breaking measurements locally disrupt this charge transport by acting as defects that create or annihilate charge with equal probability, independently of the measurement outcome, thereby preventing the development of long-range charge correlations and ruling out a charge-sharpening transition. By contrast, symmetry-preserving measurements constrain the charge to the measurement outcome and locally pin its value, enabling a charge-sharpening transition. Consequently, in the symmetry-breaking case, the charge correlation length remains finite at any measurement rate and the minimal-cut length is the only relevant length scale governing the trajectory-averaged entanglement entropy. The entanglement transition is therefore controlled solely by the connectivity between the subsystem and its complement and belongs to the classical percolation universality class, as in non-symmetric Haar-random circuits.

It would be interesting to extend our results to other classes of monitored quantum systems, such as those with deterministic or Hamiltonian unitary dynamics, particularly free-fermionic systems, for which a variety of analytical and numerical techniques have been developed~\cite{fava23,fava24,kmr23,poboiko-23,sfs25,kts25}. Other natural directions are to investigate the interplay between competing measurements~\cite{lab-21}, for example by interspersing measurements in the $Z$ and $X$ bases with different probabilities, and to consider unitary gates with symmetries beyond $U(1)$, such as discrete symmetries~\cite{khanna26,hc-22} or non-Abelian groups~\cite{magpvy-23}.
Given that the symmetry is broken, it would be particularly interesting to investigate whether, and to what extent, the entanglement asymmetry \cite{amc23,fossati24, tcdl-25, amcp-25} can provide further insight into the universal behavior identified here. A further direction is to analyze the properties of circuits with symmetric unitary gates and (symmetry-breaking) measurements away from criticality, and to compare their behavior with the universal properties of the volume-law phase identified for circuits without symmetry~\cite{deluca25, gkgdt-26,magni26}.
Finally, understanding the role of measurements in unitary evolutions with global symmetries from the perspective of the renormalization group framework~\cite{nw23} might provide a general description of their large-scale behavior.\\

\textit{Acknowledgments.---} We thank Lorenzo Piroli, Alessandro Romito, and Romain Vasseur for useful discussions. All authors acknowledge support from the European Research
Council under the Advanced Grant no. 101199196 (MOSE).

\appendix
\section{Details on the monitored SSEP model}\label{app:sm}

In this appendix, we present additional details about the one-dimensional disordered Symmetric Simple Exclusion Process (SSEP) obtained in Sec.~\ref{sec:statmech} of the main text. This model describes the contribution of the qubit sector to the average entanglement entropy of the monitored random circuit introduced in Sec.~\ref{sec:model} in the $d \to \infty$ limit. In particular, in Sec.~\ref{sec:app1} we derive our results on the typical correlation function when the measurements are performed in the $X$ eigenstates basis, while in Sec.~\ref{sec:app2} we analyze the effect of a generic symmetry-breaking measurement different from $X$.

\subsection{Maximally symmetry-breaking measurements}
\label{sec:app1}

For completeness, we summarize again the main features of the model. The configurations of the qubit sector of the random circuit are described by a set of occupation vectors $\ket{\bm{n}}$ with $n_j = 0,1$ for $j = 1,\cdots,L$. A state of the SSEP is represented by a probability vector over all microscopic configurations $\ket{P} = \sum_{\bm{n}} p_{\bm{n}}\ket{\bm{n}}$, where where $p_{\bm{n}}$ denotes the probability of the configuration $\bm{n}$. The normalization condition is $\langle\mathbbm{1}|P\rangle = 1$, where $\ket{\mathbbm{1}} = \sum_{\bm{n}} \ket{\bm{n}}$. Without monitoring, the state $\ket{P}$ is evolved with a discrete-time version of the SSEP, in particular
\begin{equation}
    \ket{P(t+1)} = \mathbb{T}_e \mathbb{T}_o \ket{P(t)},
\end{equation}
with $\mathbb{T}_{e/o} = \prod_{j \in e/o} T_{j,j+1}$, and $T_{j,j+1}$ the two-site transfer matrix, which reads
\begin{equation}\label{eq:ssep_rules}
   T = \begin{pmatrix}
        1 & 0 & 0 & 0 \\
        0 & 1/2 & 1/2 & 0 \\
        0 & 1/2 & 1/2 & 0 \\
        0 & 0 & 0 & 1 \\
    \end{pmatrix},
\end{equation}
in the basis 
$\{\ket{00},\ket{01},\ket{10},\ket{11}\}$.

As discussed in the main text, a measurement of a qubit in the basis of the $X$ eigenstates can be represented, in terms of the SSEP dynamics, as the application of the single-site operator $M$, which in the basis $\{\ket{0}, \ket{1}\}$ takes the form
\begin{equation}\label{eq:appMdef}
    M =\frac{1}{4} \begin{pmatrix}
        1 & 1 \\
        1 & 1 \\
    \end{pmatrix},
\end{equation}
for any measurement outcome. In the SSEP dynamics, measurements are implemented stochastically as follows: after the application of a single two-site update
$T_{j,j+1}$, each of the sites $j$ and $j+1$ is independently subject to a measurement with probability $p$ (the measurement rate in the circuit), in which case the corresponding local state is replaced by the uniform distribution given by Eq.~\eqref{eq:appMdef}.

Each measurement configuration $\bm{X}$ fixes the spacetime locations where the operators $M$ are inserted in the SSEP dynamics. Let $\mathbb{T}_{\bm{X}}$ denote the disordered transfer matrix evolving the state $\ket{P(0)}$ from time $0$ to time $t$ for a given $\bm{X}$, and let $N_{\bm{X}}$ be the total number of measurements contained in $\mathbb{T}_{\bm{X}}$. The expectation value of any operator $\mathcal{O}$ is then computed as $\langle \mathcal{O}(t)\rangle = 2^{N_{\bm{X}}}\bra{\mathbbm{1}} \mathcal{O} \ket{P(t)}$. The factor $2^{N_{\bm{X}}}$ arises because the measurement operators $M$ do not conserve probability. In the case of $X$-basis measurements, since the outcome probabilities are state-independent, this normalization can be absorbed into a redefinition of the measurement operator, $M \mapsto 2M$. In the following, we will therefore compute expectation values as $\langle \mathcal{O}(t)\rangle = \bra{\mathbbm{1}} \mathcal{O} \ket{P(t)}$.

Let us now study the connected two-point function of the SSEP charge in the presence of the disordered defects introduced by $M$. To evaluate this quantity, we assume, as in the main text, that the initial state is the uniform probability state $\ket{P(0)} = 2^{-L}\ket{\mathbbm{1}}$, which is left invariant by the disordered dynamics for any realization $\bm{X}$. The connected two-point correlation function $C_{\bm{X}}(z,t) \equiv\langle n(z,t) n(0,0) \rangle_c $ is defined as
\begin{equation}
    C_{\bm{X}}(z,t) = \langle n(z,t) n(0,0) \rangle - \langle n(z,t) \rangle \langle n(0,0) \rangle.
\end{equation}
We consider in the following $t \geq 0$.
The one-point function reads 
\begin{equation}
\langle n(z,t)\rangle = 2^{-L} \bra{\mathbbm{1}}n(z,t) \ket{\mathbbm{1}} = 2^{-L} \bra{\mathbbm{1}}n(z,0) \ket{\mathbbm{1}} = 1/2
\end{equation}
for all $\bm{X}$. Thus, for simplicity, we work, from now on, with the spin operator
\begin{equation}
s(z,t) \equiv 1-2 n(z,t),    
\end{equation}
since $\langle s(z, t)\rangle=0$
 for all $\bm{X}$ and 
\begin{equation}\label{eq:ssep_corr_spin}
C_{\bm{X}}(z,t) = \frac{1}{4} \langle s(z,t) s(0,0) \rangle.
\end{equation}
Given that $s(z, t)$ is diagonal in the occupation basis $\ket{\bm{n}}$, $s(z,t) \ket{\bm{n}} = (-1)^{n_z} \ket{\bm{n}}$, its time-evolution under the SSEP dynamics~\eqref{eq:ssep_rules} on sites $z$,$ z+1$ is
\begin{equation}
\label{eq:app_s_update}
    s(z,t+1) = \frac{1}{2}s(z,t) + \frac{1}{2}s(z+1,t).
\end{equation}
Equation~\eqref{eq:app_s_update} can be derived as follows. Let us consider a generic operator $\mathcal{O}$ which is necessarily diagonal in the occupation basis, i.e. $\mathcal{O}\ket{\bm{n}} = f_{\mathcal{O}}(\bm{n}) \ket{\bm{n}}$, with $f_\mathcal{O}(\bm{n}) \in \mathbb{R}$. The expectation value on a certain state $\ket{P} = \sum_{\bm{n}}p_{\bm{n}}\ket{\bm{n}}$ reads $\bra{\mathbbm{1}}\mathcal{O}\ket{P} = \sum_{\bm{s}} p_{\bm{n}} f_\mathcal{O}(\bm{n})$. Let us suppose now that the state is evolved as $\ket{P} \mapsto \mathbb{T} \ket{P}$. The time-evolved operator $\mathcal{O}^{(\mathbb{T})}$ is defined through
\begin{equation}
    \bra{\mathbbm{1}} \mathcal{O}^{(\mathbb{T})} \ket{P} = \bra{\mathbbm{1}} \mathcal{O} \mathbb{T} \ket{P}.
\end{equation}
This implies
\begin{equation}
    \sum_{\bm{n}}p_{\bm{n}} f_{\mathcal{O}^{(\mathbb{T})}}(\bm{n}) = \sum_{\bm{n}} \sum_{\bm{n}'} \mathbb{T}_{\bm{n},\bm{n}'} \,p_{\bm{n}'} f_{\mathcal{O}}(\bm{n}),
\end{equation}
where we introduced $\mathbb{T}_{\bm{n},\bm{n}'}  = \bra{\bm{n}} \mathbb{T}\ket{\bm{n}'} $. Finally, we obtain 
\begin{equation}
\label{eq:opevolv}
f_{\mathcal{O}^{(\mathbb{T})}}(\bm{n}) = \sum_{\bm{n}'} f_{\mathcal{O}}(\bm{n}') \mathbb{T}_{\bm{n}',\bm{n}},
\end{equation}
which completely determines the operator $\mathcal{O}^{(\mathbb{T})}$ since $\mathcal{O}^{(\mathbb{T})} \ket{\bm{n}} = f_{\mathcal{O}^{(\mathbb{T})}}(\bm{n}) \ket{\bm{n}}$.

Iterating the dynamical update \eqref{eq:app_s_update} using the SSEP rules~\eqref{eq:ssep_rules}, one finds that the operator $s(z,t)$ evolves as a single Brownian motion
\begin{equation}
\label{eq:sxtdyn}
    s(z,t) = \sum_y P(z,t|y,0) s(y,0),
\end{equation}
where $P(z,t|y,0)$ is the probability for a Brownian particle to propagate backward in time from position $z$ at time $t$ to position $y$ at time $t = 0$. 
As a result, inserting Eq.~\eqref{eq:sxtdyn} into the definition of the correlation function, we obtain, for the non-monitored SSEP,
\begin{equation}
    C(z,t) = \sum_{y}P(z,t|y,0) \langle s(y,0) s(0,0) \rangle.
\end{equation}
In the uniform state, $\bra{\mathbbm{1}} s(y,0) s(0,0) \ket{\mathbbm{1}} = 2^{L}\delta_{y,0}$, and
\begin{equation}
    C(z,t) = \frac{1}{4} P(z,t |0,0).
\end{equation}
This expression for the SSEP two-point function in the uniform state in terms of a single non-interacting Brownian particle is a consequence of a more general result, namely the self-duality property of SSEP connecting a $k$-point function to a $k$-particle SSEP~\cite{ss94}.

Let us now consider the effect of the measurements on the spin operator evolution. Under the action of $M$ in Eq.~\eqref{eq:appMdef} at site $z$, and using Eq.~\eqref{eq:opevolv}, we obtain 
\begin{equation}
\label{eq:sM}
    s(z,t) \xrightarrow{M} 0.
\end{equation}
The effect of measuring is therefore to annihilate the spin evolution at that site, resulting in a loss of probability conservation. Consequently, for a given measurement configuration $\bm{X}$, the correlation function~\eqref{eq:ssep_corr_spin} can be expressed as the survival probability $P_{\bm{X}}(z,t|0,0)$ of the Brownian particle propagating from $z$ at time $t$ to $0$ at time $0$,
\begin{equation}\label{eq:ssep_corr_surv}
C_{\bm{X}}(z,t) = \frac{1}{4}P_{\bm{X}}(z,t|0,0).
\end{equation}
We now average the survival probability over all possible realizations $\bm{X}$. Since the probability that a measurement is not performed is $1-p$, each Brownian trajectory contributing to $P(z,t|0,0)$ has probability $(1-p)^{2t}$ to survive. We then conclude that
\begin{equation}\label{eq:bound_survival}
\begin{split}
    \mathbb{E}_{\bm{X}}[P_{\bm{X}}(z,t|0,0)] &\leq (1-p)^{2t} P(z,t|0,0) \\
    &\leq(1-p)^{2t}.
\end{split}
\end{equation}
Applying this result in Eq.~\eqref{eq:ssep_corr_spin}, the averaged connected two-point function is bounded by 
\begin{equation}\label{eq:av_corr_ssep_bound}
\mathbb{E}_{\bm{X}}[ C_{\bm{X}}(z,t) ] \leq \frac{1}{4} (1-p)^{2t} = \frac{1}{4} e^{-t/\xi_p},
\end{equation}
which is the result in Eq.~\eqref{eq:avgC} in the main text, where $\xi^{-1}_p = -2 \log (1-p)$. 

We can further show that this average bound also implies that large correlations are exponentially unlikely in individual realizations.
We can regard $C_{\bm{X}}(z,t)$ as a positive random variable over the measurement realizations $\bm{X}$. Then, by direct application of the Markov's inequality \cite{durrett19},
\begin{equation}
\begin{split}
\label{eq:markov}
    \mathrm{Prob}_{\bm{X}}( C_{\bm{X}}(z,t) \geq e^{- t/\xi})&\leq e^{t/\xi}\mathbb{E}_{\bm{X}}[C_{\bm{X}}(z,t)] \\ &\leq e^{-(1/\xi_p-1/\xi)t},
\end{split}
\end{equation}
for any $0 < 1/\xi < 1/\xi_p$, where $\mathrm{Prob}_{\bm{X}}$ denotes the probability associated with $\bm{X}$. Equation~\eqref{eq:markov} shows that measurement realizations for which the charge correlator decays more slowly than $e^{-t/\xi}$, with $\xi>\xi_p$, occur 
with exponentially small probability. Hence, except for exponentially rare measurement realizations, the charge correlator is exponentially suppressed with 
a decay rate at least $1/\xi_p$. Moreover, since
\begin{equation}
\sum_{t = 0}^{\infty} \mathrm{Prob}_{\bm{X}}( C_{\bm{X}}(z,t) \geq e^{- t/\xi}) \leq \sum_{t = 0}^{\infty} e^{-(1/\xi_p-1/\xi)t} < \infty,
\end{equation}
we can apply the Borel-Cantelli lemma (see again Ref.~\cite{durrett19}) and prove that, \textit{with probability one} over $\bm{X}$,
\begin{equation}
    \liminf_{t\to \infty} \left(-\frac{1}{t} \log C_{\bm{X}}(z,t)\right) \geq \xi^{-1},
\end{equation}
for any fixed $\xi$ such that $0 < 1/\xi < 1/\xi_p$. This bound also implies that, with probability one,
\begin{equation}\label{eq:lim_inf}
    \liminf_{t\to \infty} \left(-\frac{1}{t} \log C_{\bm{X}}(z,t)\right) \geq \xi_p^{-1}.
\end{equation}
This result establishes that, almost surely, the charge correlator asymptotically decays exponentially in time, with an inverse correlation length bounded from below by $\xi_p^{-1}$.

Notice that the key property leading to Eq.~\eqref{eq:lim_inf} is the exponential decay~\eqref{eq:bound_survival} of the survival probability of the Brownian particle. This suggests that the same mechanism extends to higher-point functions defined on disjoint spacetime subsets with large separations.

\subsection{Measurements in a generic rotated basis}
\label{sec:app2}

Here, we discuss the case of local symmetry-breaking measurements that do not project onto the eigenstates of $X$, but rather onto the eigenstates of a generic linear superposition $\cos \theta Z + \sin \theta X$, with $0<\theta <\pi$. The Boltzmann weight of the measured bonds in the large-$d$ two-dimensional statistical model can be obtained as in Eq.~\eqref{eq:BWmeas} by computing the overlaps of $\ket{\bm{\alpha}; \sigma}$ and $\ket{\bm{\beta}; \tau}$ with the eigenstates of $\cos\theta Z + \sin\theta X$. The Boltzmann weigth leads, in the language of the SSEP model, to the following single-site operator in the basis $\{\ket{0},\ket{1}\}$
\begin{equation}\label{eq:rot_meas_ssep}
    M_{x}(\theta) = \frac{1}{4} \begin{pmatrix}
        (1+x \cos \theta)^2 & \sin^2 \theta \\
        \sin^2 \theta & (1-x \cos \theta)^2
    \end{pmatrix},
\end{equation}
corresponding to the measurement with outcome $x = \pm 1$. Observe that both the Boltzmann weight and the operator implementing the measurement in the SSEP dynamics now depend on the measurement outcome, in contrast to the case of $X$ measurements. As already stressed in the main text, this complicates the analysis, since the quantum trajectories $\bm{x}$ are not equiprobable, and the qubit entanglement entropy~\eqref{eq:Sq} must also be averaged over trajectories with Born weights $Z_0(\bm{x})$.
Another immediate consequence of taking $\theta \neq \pi/2$ is that the uniform probability state $2^{-L}\ket{\mathbbm{1}}$ is not left invariant by the monitored SSEP dynamics for individual measurement realizations.

Nonetheless, we can show that the arguments in App.~\ref{sec:app1} for the emergence of a finite correlation length continue to apply to any trajectory $\bm{x}$ when $\theta \neq \pi/2$. Let us consider a fixed realization $\bm{X}$ of the measurement locations and a fixed outcome trajectory $\bm{x}$, and analyze the spin connected two-point function for a generic initial state $\ket{P(0)}$,
\begin{equation}
\label{eq:Cc}
    C_{\bm{X},\bm{x}} (z_2,t_2;z_1,t_1) =\langle s(z_2,t_2) s(z_1,t_1)  \rangle - \langle s(z_2,t_2)\rangle \langle s(z_1,t_1) \rangle.
\end{equation}
We assume that $t_2 \geq t_1$.
The time evolution of the spin operator $s(z,t)$ under the SSEP update rules~\eqref{eq:ssep_rules} is still given by Eq.~\eqref{eq:app_s_update}. However, the measurement operator~\eqref{eq:rot_meas_ssep} with tilting angle $\theta$ and outcome $x$ maps the spin operator to 
\begin{equation}\label{eq:map_M_tilted}
    s(z,t)  \xrightarrow{M_{x}(\theta)} \cos^{2}\theta \,s(z,t) + x \cos \theta \, \mathbb{I},
\end{equation}
where we used again Eq.~\eqref{eq:opevolv} to derive it.
Unlike in the $X$ measurement basis (cf. Eq.~\eqref{eq:sM}), the spin operator is not completely annihilated by the measurement~\eqref{eq:rot_meas_ssep}, but instead survives with a $\cos^2\theta$ prefactor. The full time evolution is thus given by
\begin{equation}
\label{eq:st_theta}
    s(z,t) = \sum_{y} P_{\bm{X}}(z,t|y,t_1) s(y,t_1) + R_{\bm{x}} \cos\theta \,\mathbb{I},
\end{equation}
where $P_{\bm{X}}(z,t|y,t_1)$
is again the survival probability of a Brownian particle propagating backward from $z$ at time $t$ to $y$ at time $t_1$, as in the case of $X$ measurements, and 
$R_{\bm{x}} = \sum_j x_j$ is the sum of the outcomes of all measurements acting on the spin operator from time $t_1$ (included) to time $t$ (excluded).

Plugging Eq.~\eqref{eq:st_theta} in the connected correlation function \eqref{eq:Cc}, the contribution from the term $R_{\bm{x}} \cos\theta \,\mathbb{I}$ in Eq.~\eqref{eq:st_theta} cancels exactly and we obtain
\begin{equation}
    C_{\bm{X},\bm{x}} (z_2,t_2;z_1,t_1) = \sum_y P_{\bm{X}}(z_2,t_2|y,t_1) \,C_{\bm{X},\bm{x}} (y,t_1;z_1,t_1).
\end{equation}
Using now $0<C_{\bm{X},\bm{x}}<1$, we find the simple bound
\begin{equation}
    C_{\bm{X},\bm{x}} (z_2,t_2;z_1,t_1) \leq \sum_y P_{\bm{X}}(z_2,t_2|y,t_1).
\end{equation}
The average correlation over the measurement locations $\bm{X}$ can be bounded as in the case of the $X$ measurement, except that the survival probability per measurement is now modified according to Eq.~\eqref{eq:map_M_tilted}. Each Brownian trajectory contributing to $P(z, t|y, t_1)$ has now probability to survive $(1-p+p\cos^2\theta)^{2t}$, where $1-p$ is the probability that no measurement is performed in a site and $p\cos^2\theta$ is the probability that the operator survives the measurement, as described in Eq.~\eqref{eq:map_M_tilted}. Therefore, we have 
\begin{equation}
\label{eq:avgCtheta}
\begin{split}
    \mathbb{E}_{\bm{X}}[C_{\bm{X},\bm{x}} (z_2,t_2;z_1,t_1)] &\leq (1- p + p \cos^2 \theta)^{2(t_2-t_1)}\\
    &\leq e^{-t/\xi_p(\theta)},
\end{split}
\end{equation}
for any trajectory $\bm{x}$ and any initial state $\ket{P(0)}$. Here, we have defined the finite correlation length $\xi_p(\theta)^{-1} = -2 \log(1-p \sin^2 \theta)$.

Using the bound in Eq.~\eqref{eq:avgCtheta}, we can prove that all of the results derived in the previous section apply also for any finite $0<\theta<\pi/2$ with the substitution $\xi_p \mapsto \xi_p(\theta)$. In particular, Eqs.~\eqref{eq:markov} and~\eqref{eq:lim_inf} hold, implying that, with probability one over the measurement realizations $\bm{X}$, the charge correlator~\eqref{eq:Cc} decays exponentially in time with an inverse correlation length bounded from below by $\xi_p(\theta)^{-1}$, for any trajectory $\bm{x}$. 

\section{Random measurement basis model}
\label{app:rndbas}

In this section, we analyze a different measurement protocol for the monitored $U(1)$-symmetric Haar circuit and prove that, in the large-$d$ limit, it is exactly equivalent to the statistical model associated with $X$ measurements. Instead of performing local projective measurements in a fixed basis of the local Hilbert space, we consider an independently chosen random basis of the local space $\mathbb{C}^2 \otimes \mathbb{C}^d$ for each measurement. We realize this measurement scheme in our circuit model by rotating the projectors of the qubit, initially in the $Z$ eigenbasis, and of the qudit, in an arbitrary basis, with a one-site Haar-random unitary acting on the full local Hilbert space. Notice that, because of the invariance of the Haar measure, the choice of the initial basis is completely irrelevant. Therefore, in each measurement location, we apply the projector $\Pi_{\alpha,m}^{(u)} = u \ket{\alpha,m} \bra{\alpha,m}u^{\dagger}$, where $\alpha = \pm 1$ labels the outcome of the $Z$ measurement in the qubit, $m = 1,\cdots,d$ labels the outcome in the fixed basis of the qudit, and $u \in U(2d)$ is a one-site Haar-random unitary matrix acting on $\mathbb{C}^2\otimes \mathbb{C}^d$.

As explained in Sec.~\ref{sec:statmech} of the main text, to compute the trajectory-averaged entanglement entropy we first replicate the circuit $Q$ times and then perform the average over gate realizations, which now also includes the independently drawn one-site unitaries $u$ that rotate the measurement basis. The Boltzmann weight associated to a measured bond is given now by (cf.~\eqref{eq:BWmeas})
\begin{equation}
\label{eq:rndBWmeas}
\mathbb{E}_{u}[\langle\bm{\alpha};\sigma|\Pi^{(u)}_{\gamma,m}\rangle \langle\Pi^{(u)}_{\gamma,m} |\bm{\beta};\tau \rangle],
\end{equation}
where, similarly as in the main text, we have introduced $\ket{\Pi^{(u)}_{\gamma,x}} = (u\otimes u^*)^{\otimes Q} \,(\ket{\gamma,m} \otimes \ket{\gamma,m})^{\otimes Q}$. We rearrange the two contractions in Eq.~\eqref{eq:rndBWmeas} as a single contraction in the $2Q$-fold replicated local Hilbert space as 
\begin{equation}
\label{eq:rndBW1}
    \left(\bra{\Pi_{\gamma,m}} \otimes \bra{\Pi_{\gamma,m}}\right)  \mathbb{E}_{u}[(u \otimes u^*)^{\otimes 2 Q}] \left(\ket{\bm{\alpha};\sigma} \otimes \ket{\bm{\beta},\tau}\right).
\end{equation}
The expectation value over the replicated single-site unitaries can be obtained applying Weingarten calculus~\cite{weingarten-78, bb-96, cs-06}. In the large-$d$ limit, the result is
\begin{equation}
\label{eq:diagWeing}
    \mathbb{E}_{u}[(u \otimes u^*)^{\otimes 2 Q}] \simeq\frac{1}{(2d)^{2Q}} \sum_{\rho \in S_{2Q}} \ket{\rho} \bra{\rho}.
\end{equation}
The permutation states appearing in Eq.~\eqref{eq:diagWeing} can be expressed in terms of the fixed-qubit-charge states $\{\ket{\bm{\alpha};\sigma}\}$, defined in Eq.~\eqref{eq:permstates}, as
\begin{equation}
    \ket{\rho} = \sum_{\bm{\mu}}\ket{\bm{\mu};\rho}.
\end{equation}
Plugging Eq.~\eqref{eq:diagWeing} into Eq.~\eqref{eq:rndBW1}, we obtain
\begin{equation}
\label{eq:rndBW2}
    \mathbb{E}_{u}[\langle\bm{\alpha};\sigma|\Pi^{(u)}_{\gamma,m}\rangle \langle\Pi^{(u)}_{\gamma,m} |\bm{\beta};\tau \rangle] = \frac{1}{(2d)^{2 Q}} \sum_{\rho \in S_{2Q}} \sum_{\bm{\mu},\bm{\nu}} \left(\bra{\Pi_{\gamma,m}} \otimes \bra{\Pi_{\gamma,m}}\right) \cdot \ket{\bm{\mu};\rho}\, \bra{\bm{\nu};\rho} \cdot\left(\ket{\bm{\alpha};\sigma} \otimes \ket{\bm{\beta},\tau}\right).
\end{equation}
In the large-$d$ limit, the rightmost contraction simplifies significantly because the permutation $\rho$ over $2Q$ replicas is constrained to coincide with the permutation obtained by composing $\sigma$ on the first $Q$ replicas and $\tau$ on the last $Q$ replicas, which we denote by $(\sigma)(\tau)$.
Moreover, the charge configuration $\bm{\nu}$ in the $2Q$ replicas is completely fixed by the charge configurations $\bm{\alpha}$ and $\bm{\beta}$ in the first and last $Q$ replicas, respectively, so that we have
\begin{equation}
    \bra{\bm{\nu};\rho} \cdot\left(\ket{\bm{\alpha};\sigma} \otimes \ket{\bm{\beta},\tau}\right) \simeq d^{2 Q} \delta_{\rho,(\sigma)(\tau)} \delta_{\bm{\nu},{(\bm{\alpha})(\bm{\beta})}}.
\end{equation}
Applying this result in Eq.~\eqref{eq:rndBW2}, we obtain
\begin{equation}
\label{eq:rndBWfin}
\begin{split}
    \mathbb{E}_{u}[\langle\bm{\alpha};\sigma|\Pi^{(u)}_{\gamma,m}\rangle \langle\Pi^{(u)}_{\gamma,m} |\bm{\beta};\tau \rangle] &= \frac{1}{4^Q} \sum_{\bm{\mu}} \left(\bra{\Pi_{\gamma,m}} \otimes \bra{\Pi_{\gamma,m}}\right) \cdot \ket{\bm{\mu}; (\sigma)(\tau)} \\
    &= \frac{1}{4^Q} \sum_{\bm{\mu}_1,\bm{\mu}_2} \langle \Pi_{\gamma,m}| \bm{\mu}_1;\sigma\rangle \langle \Pi_{\gamma,m}| \bm{\mu}_2;\tau\rangle \\
    &= \frac{1}{4^{Q}}.
\end{split}
\end{equation}
In the last equality, we have used the fact that the contraction with the projector state equals $1$ only for charge configurations in which the charge is the same in each replica and matches the measurement outcome. The final result in Eq.~\eqref{eq:rndBWfin} coincides with that obtained in Eq.~\eqref{eq:BWmeas} of the main text for measurements in the $X$ eigenbasis.

\section{Details on the finite-size numerical analysis}
\label{app:num}

\begin{figure*}[t]
\includegraphics[width=0.96\textwidth]{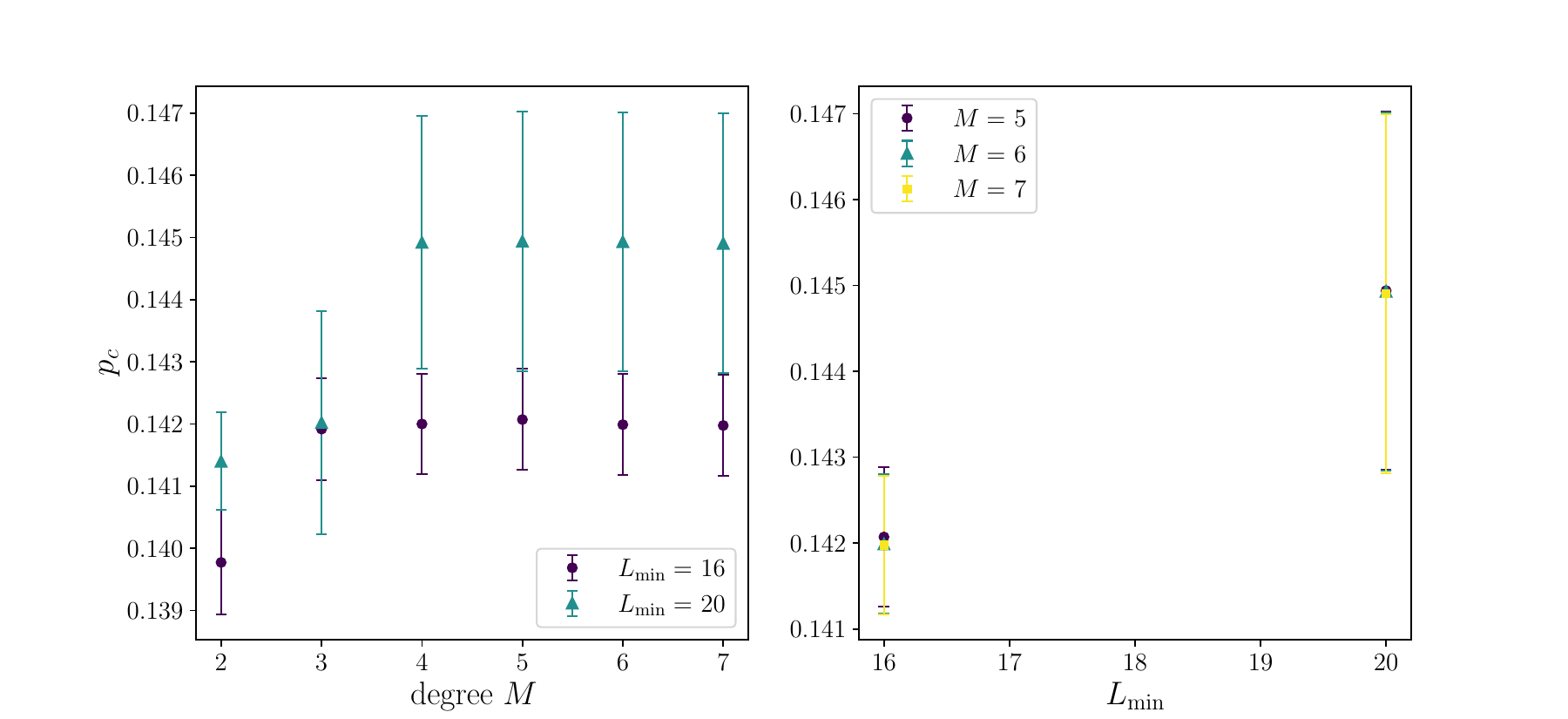}
\includegraphics[width=0.96\textwidth]{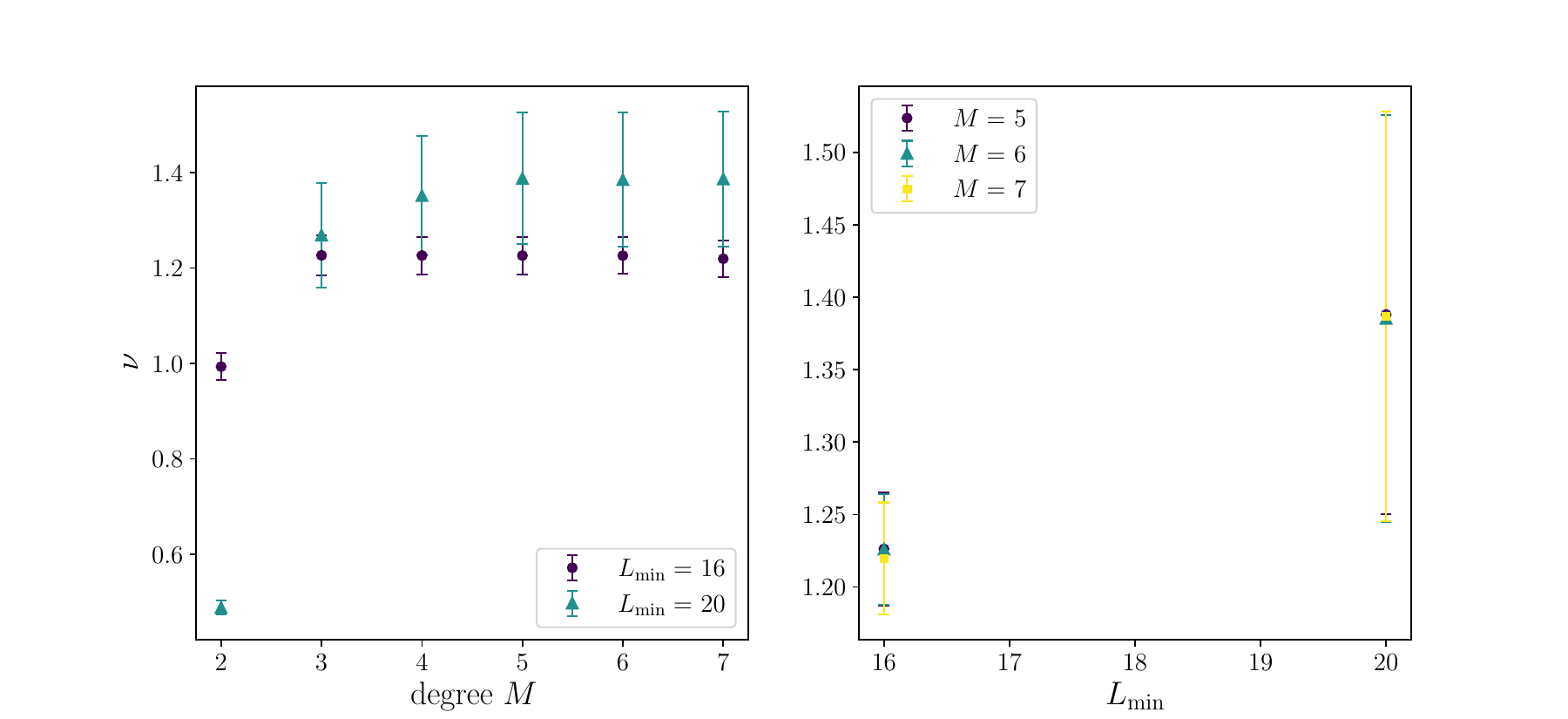}
\centering
\caption{Estimates of the critical measurement rate $p_c$ (upper panels) and critical exponent $\nu$ (lower panels) for the $X$-monitored $U(1)$ Haar-random circuit, obtained from a non-linear fit of the function in Eq.~\eqref{eq:polyfit} to numerical simulation data. Left panels: we vary the degree $M$ of the fitting function for different minimum system sizes $L_{\rm min}$ included in the dataset. Right panels: the same data are shown explicitly as a function of $L_{\rm min}$.}
\label{fig:pcHaar} 
\end{figure*}

In this Appendix, we describe the numerical procedures used to perform the finite-size scaling analysis of the MIPT in the monitored circuits studied in Secs.~\ref{sec:qubitnum} and \ref{sec:stab}. 

\begin{figure*}[t]
\includegraphics[width=0.6\textwidth]{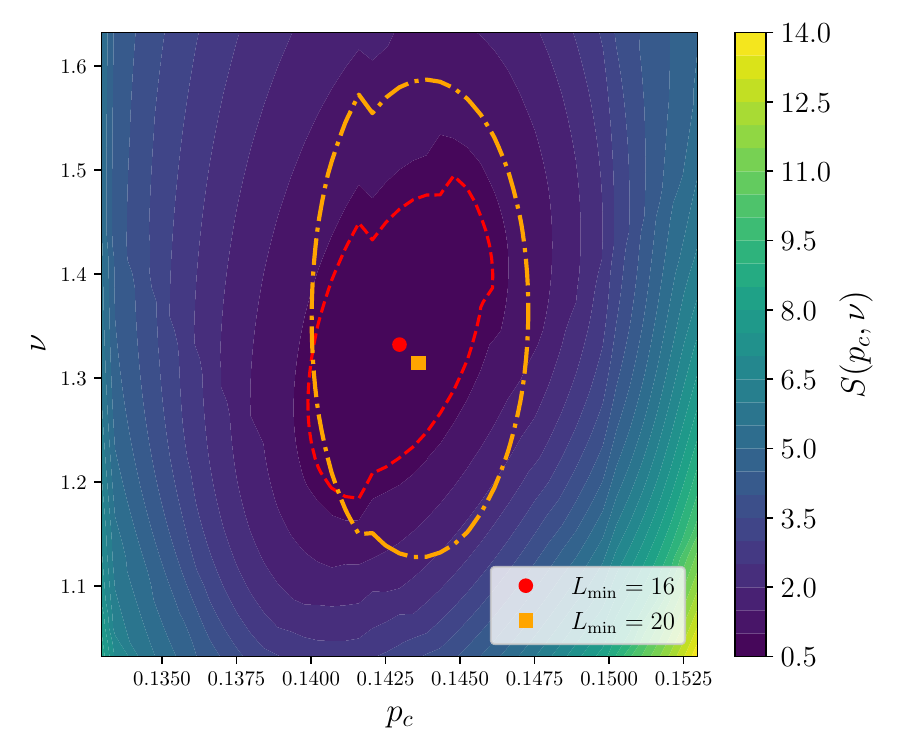}
\centering
\caption{Heatmap of the cost function $S(p_c,\nu)$ defined in Eq.~\eqref{eq:costf}, for the numerical simulation data of the $X$-monitored $U(1)$ Haar-random circuit. The symbols indicate the location of the  global minimum by varying the minimum system size $L_{\min}$ included in the dataset: red circle for $L_{\min} = 16$ and orange square for $L_{\min} = 20$. The dashed lines delimit the region $S(p_c,\nu) = 1.3 \, S(p_c^*,\nu^*)$, which provides an estimate of the error for the values $p_c^*$ and $\nu^*$ that minimize the cost function.}
\label{fig:objHaar} 
\end{figure*}

The scaling ansatz for the average tripartite mutual information $\mathbb{E}[I_3]$, neglecting the subleading finite-size corrections coming from RG irrelevant operators, is (Eq.~\eqref{eq:fssI3})
\begin{equation}
\mathbb{E}[I_{3}] = \mathcal{I}\left[(p-p_c)L^{1/\nu}\right]. 
\end{equation}
We have used two different approaches to determine the optimal $p_c$ and $\nu$. The first method consists in performing a polynomial approximation of the scaling function $\mathcal{I}$ in terms of the variable $x \equiv (p-p_c)L^{1/\nu}$,
\begin{equation}
\label{eq:polyfit}
    \mathbb{E}[I_3] \approx \sum_{j = 0}^M c_j \,x^j.
\end{equation}
Using this polynomial, we carry out a non-linear fit of the numerically obtained data for system sizes $L$ larger than a chosen minimum $L_{\min}$. The fit provides estimates for $p_c$, $\nu$, and the coefficients $\{c_j\}_{j=0,\cdots,M}$. We then repeat the fit for different values of the polynomial degree $M$ and test the stability of the resulting parameter estimates. We further investigate the dependence of the fit results on the minimum system size $L_{{\rm min}}$ by varying it. By doing so, we are able to estimate the systematic error associated with the finite-size scaling ansatz arising from finite-size effects.

The second method used to analyze the same data is based on Ref.~\cite{ki93} and has also been employed in the context of monitored circuits in Ref.~\cite{zabalo20}. It relies on the minimization of a cost function $S(p_c,\nu)$. For given values $p_i$ and $L_i$ of the measurement rate and system size, we denote $x_i = (p_i-p_c)L_i^{1/\nu}$, $y_i = \mathbb{E}[I_3]$, the average value of $I_3$ for the corresponding $(p_i,L_i)$, computed over the circuit realizations sampled in the numerical simulations, and $d_i = \sigma_{y_i}$, the statistical uncertainty associated with $y_i$. The resulting dataset $\{x_i,y_i,d_i\}_{i=1}^{N_s}$ is ordered such that $x_i<x_{i+1}$. The cost function $S(p_c,\nu)$ is defined as
\begin{equation}
\label{eq:costf}
    S(p_c,\nu) = \frac{1}{N_s-2} \sum_{i=2}^{N_s-1} w(x_i,y_i,d_i | x_{i-1},y_{i-1},d_{i-1};x_{i+1},y_{i+1},d_{i+1}),
\end{equation}
where the quantity $w(x,y,d|x',y',d' ; x'',y'',d'')$ is
\begin{equation}
    w \equiv \left( \frac{y-\bar{y}}{\Delta(y-\bar{y})} \right)^2,
\end{equation}
with
\begin{equation}
    \bar{y} \equiv \frac{(x''-x)y' - (x'-x)y''}{x''-x},
\end{equation}
and
\begin{equation}
    (\Delta(y-y'))^2 \equiv d^2 + \left(\frac{x''-x}{x''-x'} d'\right)^2 + \left(\frac{x'-x}{x''-x'} d''\right)^2.
\end{equation}
The cost function~\eqref{eq:costf} quantifies how closely each point $(x_i,y_i)$ aligns with the linear interpolation defined by its adjacent neighbors $(x_{i-1},y_{i-1})$ and 
$(x_{i+1},y_{i+1})$. For an ideal collapse, the minimal value of $S(p_c, \nu)$ should be close to $1$.
We then estimate the optimal $p_c^*$ and $\nu^*$ by locating the global minimum of $S(p_c,\nu)$. The error of these optimal values is determined by identifying the region of parameters where $S(p_c,\nu) \leq 1.3 \,S(p_c^*,\nu^*)$~\cite{zabalo20}.

We have verified that the estimates obtained from the two methods are compatible. In the following, we report this analysis for the two circuits considered in the main text.

\begin{figure*}[t]
\includegraphics[width=0.96\textwidth]{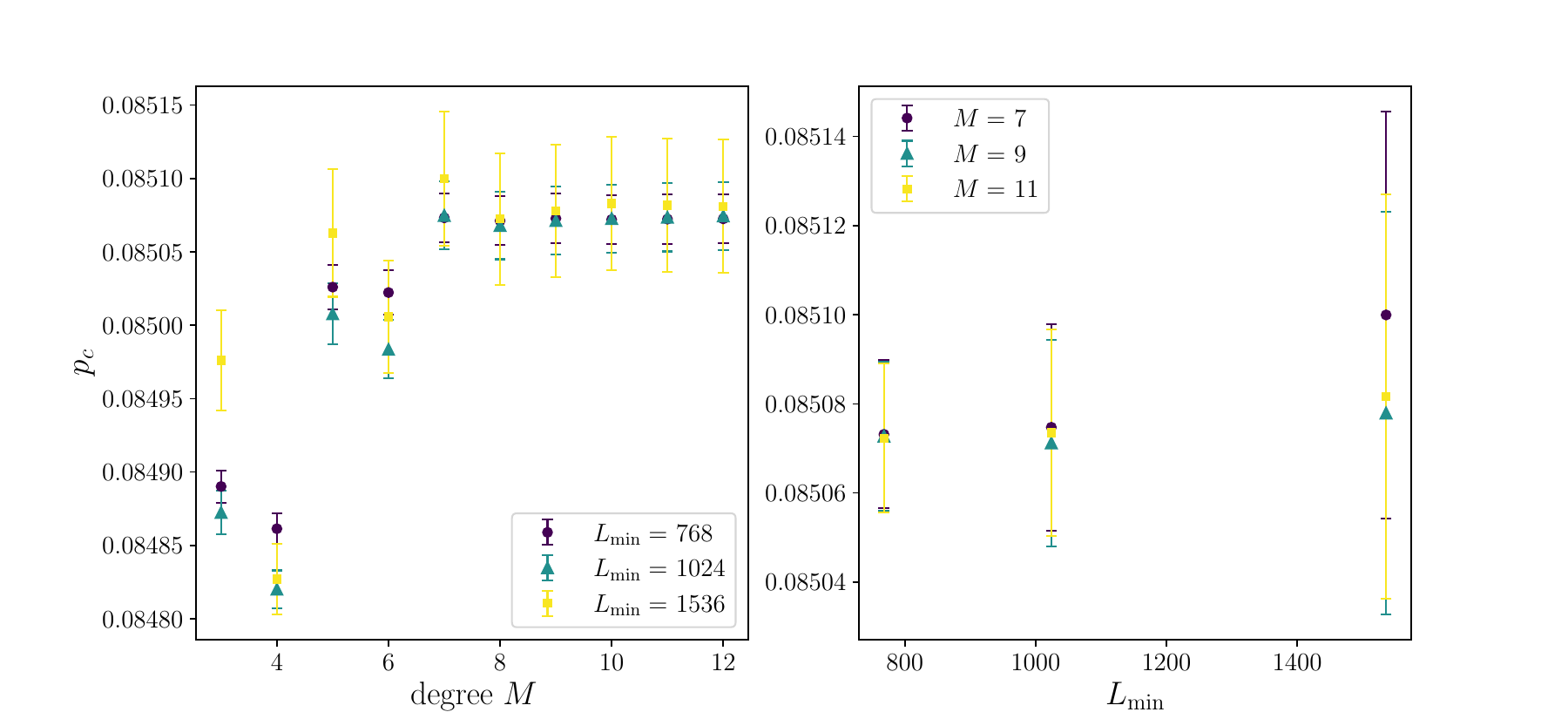}
\includegraphics[width=0.96\textwidth]{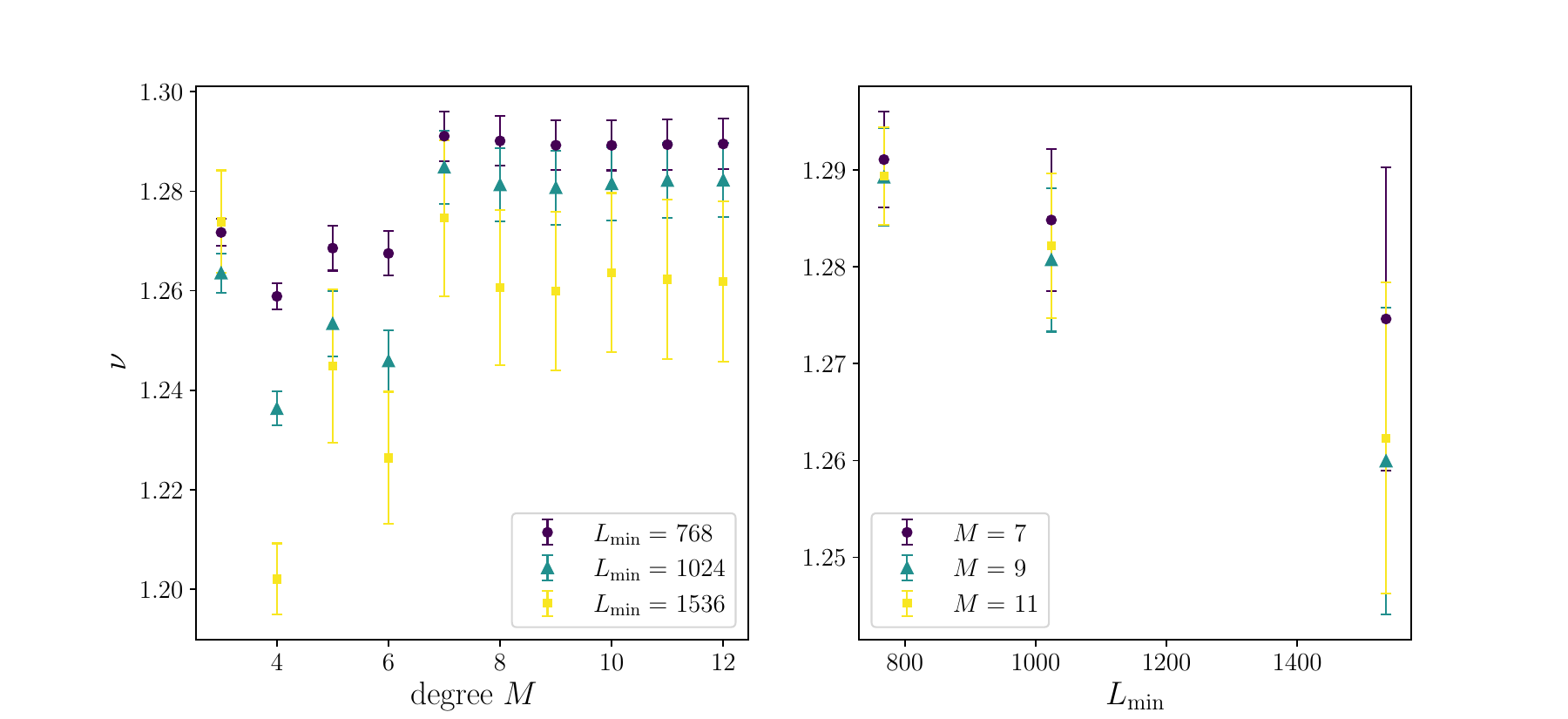}
\centering
\caption{Same as Fig.~\ref{fig:pcHaar}, but for the $X$-monitored $U(1)$ stabilizer circuit.}
\label{fig:pcCliff} 
\end{figure*}

\subsection{Haar circuit}

We discuss here the details of the numerical analysis for the monitored Haar circuit discussed in Sec.~\ref{sec:qubitnum}. For this model, we restrict the analysis to system sizes $L = 16,20,24$, for which we do not observe a finite-size drift of the crossing point (see Fig.~\ref{fig:I3n1} of the main text). We first present, in Fig.~\ref{fig:pcHaar}, the results for the critical measurement rate $p_c$ (upper panels) and the correlation-length exponent $\nu$ (lower panels) obtained from the non-linear fit of Eq.~\eqref{eq:polyfit}. We show the dependence of the fit estimates on the polynomial degree $M$ (left panels) and on the minimum system size $L_{\rm min}$ included in the fit (right panels). We find that both the estimated $\nu$ and $p_c$ are stable for $M\geq 4$ for the two $L_{\rm min}$ considered, whereas they exhibit a noticeable dependence on $L_{\rm{min}}$, which constitutes the dominant source of error in our results.

In Fig.~\ref{fig:objHaar}, we show a heat map of the cost function $S(p_c,\nu)$ of Eq.~\eqref{eq:costf}, 
computed using the data shown in Fig.~\ref{fig:I3n1}. The symbols indicate the location of its minimum for $L_{\rm{min}}=16$ and $20$. 
The dashed and dot-dashed curves delimit the uncertainty region defined by $S(p_c,\nu)\leq 1.3S(p_c^*,\nu^*)$ for 
$L_{\rm{min}}=16$ and $20$, respectively. This approach yields more stable estimates of $p_c$
and $\nu$ than the nonlinear fit analysis. Nevertheless, the two procedures produce compatible results.

\subsection{Stabilizer circuit}

In this subsection, we discuss the details of the numerical analysis for the monitored stabilizer circuit defined in Sec.~\ref{sec:stab}.
In this case, we consider systems of size $L\geq 768$, for which no drift of the crossing point is observed in the left panel Fig.~\ref{fig:I3Cliff}.
In Fig.~\ref{fig:pcCliff}, we present the results of the non-linear fit for the critical exponent $\nu$ (lower panels) and the measurement rate $p_c$ (upper panels) as a function of the polynomial degree $M$ (left panels) and the minimal system size $L_{{\rm min}}$ (right panels). As occurred in the Haar circuit, both $\nu$ and $p_c$  stabilize as $M$ increases for all $L_{{\rm min}}$ considered. On the other hand, the estimate of the critical point $p_c$ appears stable upon varying $L_{\rm min}$, while the critical exponent $\nu$ instead exhibits a mild dependence on $L_{\rm min}$, which dominates its uncertainty.

\begin{figure*}[t]
\includegraphics[width=0.6\textwidth]{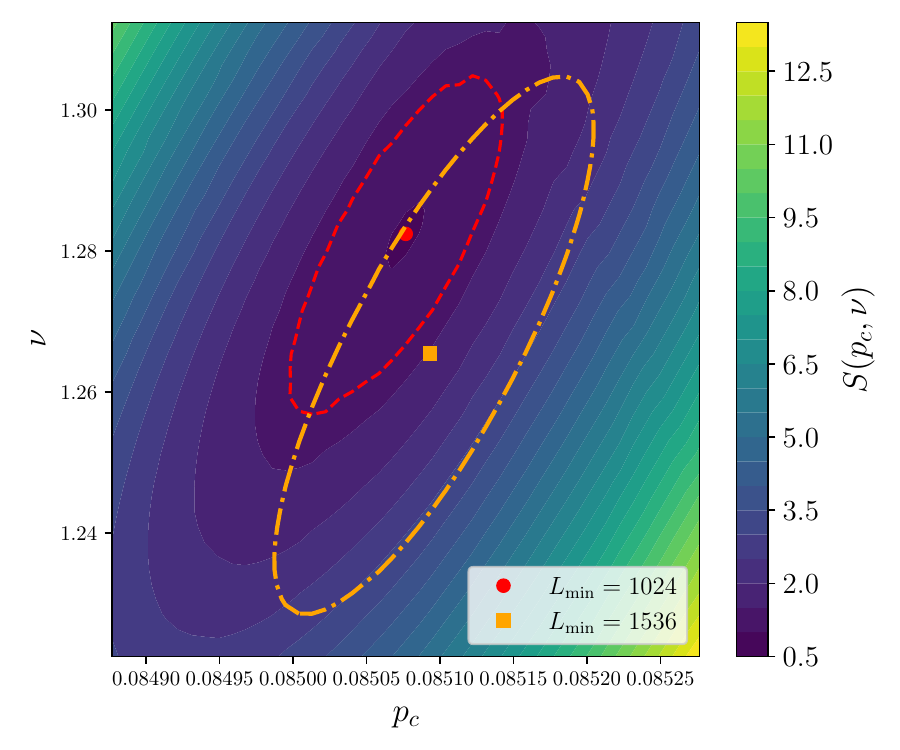}
\centering
\caption{Same as Fig.~\ref{fig:objHaar}, but for the $X$-monitored $U(1)$ stabilizer circuit.}
\label{fig:objCliff} 
\end{figure*}

Finally, we show in  Fig.~\ref{fig:objCliff} the heat map of the cost function~\eqref{eq:costf} for the stabilizer circuit using the data of the left panel of Fig.~\ref{fig:I3Cliff}. The symbols are the location of its minimum when considering $L_{\rm min}=1024$ and $L_{\rm min}=1536$. The dashed and dot-dashed curves delimit the region $S(p_c,\nu)\leq 1.3S(p_c^*,\nu^*)$ that determines the error of 
the estimates for $p_c^*$ and $\nu^*$. We observe that, as in the non-linear fit analysis, the estimate of $p_c^*$ is more stable than that of $\nu^*$ when varying $L_{\rm min}$.

\end{document}